\documentclass[12pt,superscriptaddress,floatfix,preprintnumbers,nofootinbib]{revtex4}              
\usepackage{graphicx}
\usepackage{amssymb}
\usepackage{rotating}
\usepackage{epsfig}
\usepackage[lofdepth,lotdepth,caption=false]{subfig}
\usepackage{color}
\def\be{\begin{equation}}
\def\ee{\end{equation}}
\def\bea{\begin{eqnarray}}
\def\eea{\end{eqnarray}}

\begin{document}

\title{Restricting the LSND and MiniBooNE sterile neutrinos 
with the IceCube atmospheric neutrino data}
\author{Arman~Esmaili} 
\affiliation{Instituto de Fisica Gleb Wataghin - UNICAMP, 13083-859, Campinas, SP, Brazil}
\author{Alexei~Yu.~Smirnov} 
\affiliation{Abdus Salam International Centre for Theoretical Physics, ICTP, I-34010, Trieste, Italy}

\begin{abstract}

We study oscillations of the high energy atmospheric neutrinos in the Earth into sterile neutrinos with the eV-scale mass. The MSW resonance and parametric enhancement of the $\bar{\nu}_\mu\to\bar{\nu}_s$ oscillations lead to distortion of the zenith angle distribution of the muon-track events which can be observed by IceCube. Due to matter effect, the IceCube signal depends not only on the mixing element $U_{\mu 4}$ relevant for LSND and MiniBooNE but also on $U_{\tau 4}$ and the CP-violating phase $\delta_{24}$. We show that the case with $U_{\tau 4} = \delta_{24} = 0$ leads to the weakest IceCube signal and therefore should be used to bound $U_{\mu 4}$. We compute the zenith angle distributions of the $\nu_\mu-$events for different energy intervals in the range (0.1 - 10) TeV and find that inclusion of the energy information (binning in energy) improves the sensitivity to $\nu_s$ drastically. We estimate that with already collected (during 3 - 4 years) IceCube statistics the bound $|U_{\mu 4}|^2 < 0.01$ ($99\%$ C.L.) can be established and the mixing required by LSND and MiniBooNE can be excluded at $(4 - 6) \sigma$ confidence level. 

\end{abstract}


\maketitle

\section{Introduction}
\label{sec:intro}

Existence of sterile neutrinos with mass $\sim\mathcal{O}(1)$~eV and with mixing required by the LSND~\cite{Aguilar:2001ty} and MiniBooNE~\cite{AguilarArevalo:2012va} anomalies is not a small perturbation of the standard $3\nu-$scheme. It leads to rich phenomenology and far going consequences for theory. Therefore checks of the existence of these neutrinos become one of the main objectives of the neutrino physics. Recall that the $\nu_\mu\to\nu_e$ and $\bar{\nu}_\mu\to\bar{\nu}_e$ oscillations interpretation of the LSND and MiniBooNE results implies non-vanishing admixtures of the $\nu_e$ and $\nu_\mu$ in the 4th mass eigenstate, quantified respectively by mixing elements $U_{e4}$ and $U_{\mu4}$, and the oscillation depth is given by $4|{U_{e4}U_{\mu4}}|^2$. 

The present experimental situation is rather controversial.

(i) Interpretation of the LSND/MiniBooNE results in terms of oscillations with sterile neutrinos is not very convincing in view of uncertainties in the cross-sections, energy scale calibration as well as  backgrounds~\cite{HARPCDP:2011aa}. Moreover, fit of the energy spectra of excess of events in MiniBooNE with one sterile neutrino is rather poor. Good fit can be obtained in the presence of second sterile neutrino with large mass $\Delta m^2_{51}\sim 20$~eV$^2$~\cite{Conrad:2013mka}. However, the latter is in serious conflict with Cosmology and laboratory observations. 

(ii) There is strong tension between the appearance (LSND/MiniBooNE) and disappearance (short baseline) experimental results within both 3+1 and 3+2 schemes~\cite{Kopp:2013vaa}. Furthermore, the $\nu_\mu-$disappearance has not been observed by MiniBooNE itself, thus leading to constraints on $U_{\mu4}$~\cite{AguilarArevalo:2009yj}. Strong bound follows from the neutral current interaction measurements in near and far MINOS detectors~\cite{Adamson:2011ku}. A combination of the negative results of CDHS~\cite{Dydak:1983zq}, MiniBooNE and MINOS on the $\nu_\mu-$disappearance puts the strongest limit on $U_{\mu 4}$. 

(iii) The $\nu_e-$appearance signal has not been observed in OPERA~\cite{Agafonova:2013xsk} and ICARUS~\cite{Antonello:2012pq} experiments. This directly excludes the low $\Delta m^2_{41}$ part of the LSND/MiniBooNE region.

(iv) The reactor anomaly  -- disappearance of the $\bar{\nu}_e-$flux from reactors~\cite{Mueller:2011nm}, is in favor of the eV mass scale sterile neutrinos. Still the claimed deficit of the signal can be due to underestimated uncertainties in the antineutrino flux calculations~\cite{Hayes:2013wra}.

In this connection large number of new experiments has been proposed to test existence of sterile neutrinos (see~\cite{new} and references therein, see also~{\cite{new1}) and some of them can be realized in the next 5 - 10 years. At the same time, study of the atmospheric neutrinos by IceCube with {\it already collected statistics} can contribute in substantial or even decisive way to resolution of the LSND/MiniBooNE anomaly.  

Indeed, for $\Delta m_{41}^2 \sim (1 - 10)$~eV$^2$ implied by LSND, oscillations of neutrinos with energies $(1 - 10)$~TeV in the matter of the Earth will undergo the MSW resonance enhancement~\cite{nikolaidis,yasuda,Nunokawa:2003ep}. Cosmic neutrinos would be affected by these oscillations~\cite{nikolaidis,yasuda}. The resonantly enhanced $\nu_\mu - \nu_s$ oscillations lead to partial disappearance of the $\nu_\mu$ (or $\bar{\nu}_\mu$) flux, and consequently, to distortion of the energy and zenith angle distributions of the muon neutrino charged current events~\cite{Nunokawa:2003ep}. This can be explored in IceCube using the atmospheric neutrino flux~\cite{Nunokawa:2003ep}. Actually, the resonance enhancement of the $\bar{\nu}_\mu - \bar{\nu}_s$ oscillations occurs for the mantle crossing trajectories  with the MSW resonance peak at $E_\nu\sim 4~{\rm TeV}~(\Delta m^2_{41}/{\rm eV}^2)$~\cite{Nunokawa:2003ep}. For the core crossing trajectories the parametric enhancement of oscillations takes place~\cite{parametric} at about 2 times lower energies. The original consideration for single sterile neutrino was generalized later to the case of two sterile neutrinos~\cite{Choubey:2007ji}.

The first results from AMANDA and IceCube-40 experiments motivated further detailed studies of these oscillation effects~\cite{Razzaque:2012tp,Razzaque:2011ab,Barger:2011rc,Esmaili:2012nz,Esmaili:2013cja}. In~\cite{Razzaque:2011ab} the effects of sterile neutrinos have been studied in different mixing schemes. It was observed that in the case of $\nu_s - \nu_\mu$ mixing the effects, and consequently, bounds on the mixing angle become weaker. The effect of $\nu_s$ can also be observed at lower energies, $E_\nu < 100$ GeV, in the DeepCore experiment~\cite{Razzaque:2012tp}. The analysis of DeepCore has been extended to $3 + 2$ and $1 + 3 + 1$ models with two sterile neutrinos in~\cite{Barger:2011rc}. In~\cite{Esmaili:2012nz} by scanning the parameter space of the  $3+1$ model, which includes $\{U_{\mu 4},  U_{\tau 4}, \Delta m_{41}^2\}$, a mixing scheme independent exclusion region has been found in ($U_{\mu 4},\Delta m_{41}^2$) plane from the IceCube-40 data. The bound on mixing in~\cite{Esmaili:2012nz} does not exclude the favored region by MiniBooNE/LSND. In~\cite{Esmaili:2013cja} the potential of cascade  IceCube events  in constraining sterile neutrinos has been studied.

Recently results from IceCube experiment collected during the period from May 2010 to May 2011 have been published~\cite{Aartsen:2013jza} (the so-called IceCube-79 data). The zenith angle distribution of events is in very good agreement with the standard $3\nu-$oscillations. The statistical errors in each of 10 zenith angle bins are below $(2 - 3) \%$. The systematic errors are still rather large, however small spread of the experimental points within the statistical errors indicate that systematic errors are correlated. The expected effect from sterile neutrinos is $(5 - 10) \%$ in several vertical bins~\cite{Razzaque:2012tp,Barger:2011rc,Esmaili:2012nz}. This indicates that effects of sterile neutrinos, if exist,  should be small. 

After IceCube-79 the IceCube-86 with larger effective area is taking data. Presently the IceCube-86 exposure is at least 4 times larger than the IceCube-79 one and therefore statistical error is reduced by factor 2. Also systematic error is expected to be smaller. With this one can perform critical test of  existence of the LSND/MiniBooNE sterile neutrinos. 

To exclude the LSND and MiniBooNE sterile neutrino one needs to explore effects in whole range of relevant parameters including mixing angles and CP-violating phases. In general the $4\nu-$mixing  is described by  6 mixing angles and 3 CP-violating phases. Therefore complete scanning of the parameter space is very cumbersome. Fortunately, the effect in IceCube depends appreciably only on few parameters. In this paper we identify these relevant parameters and find their values which minimize the sterile neutrino effects in IceCube. We explore ways to improve sensitivity of IceCube to sterile neutrinos using information about energies of events (energy binning).  

The paper is organized as follows. In Sec.~\ref{sec:prob} we consider in detail dependence of the $\nu_\mu$ oscillation probabilities relevant for IceCube on mixing scheme, and in particular on the $\nu_\tau-$mixing. We evaluate also effect of CP-violating phases. We show that flavor mixing scheme with $U_{\tau 4} = \delta_{\tau 4} = 0$ provides the weakest IceCube signal and therefore should be used to exclude the sterile neutrino interpretation of the LSND and MiniBooNE results. In Sec.~\ref{sec:dis} the zenith angle distributions of events in IceCube for different energy ranges are computed. Also we present the energy distributions of events smeared with the neutrino energy reconstruction function. In Sec.~\ref{sec:sen} we perform simple statistical analysis evaluating sensitivity of IceCube to the sterile neutrino mixing. We explore how the energy information will improve the sensitivity. Conclusions are presented in Sec.~\ref{sec:conc}.

\section{LSND/MiniBooNE and IceCube signals}
\label{sec:prob} 

\subsection{Generalities}

We will consider mixing of 4 neutrinos $\nu_f^T \equiv (\nu_e, \nu_\mu, \nu_\tau, \nu_s)$:  $\nu_f = \mathbf{U_4} \nu_{\rm mass}$, where $\nu_{\rm mass}^T \equiv (\nu_1,  \nu_2, \nu_3,  \nu_4)$. The unitary mixing matrix $\mathbf{U_4}$ is usually parametrized as 
\begin{equation}
\label{eq:U}
\mathbf{U_4}= \mathbf{R^{34}}(\theta_{34})\mathbf{R^{24}_{\delta}} (\theta_{24}, \delta_{24}) 
\mathbf{R^{14}_{\delta}} (\theta_{14}, \delta_{14})
\mathbf{R^{23}}(\theta_{23})\mathbf{R^{13}_{\delta}} (\theta_{13}, \delta_{13})\mathbf{R^{12}}(\theta_{12})~,
\end{equation}
where $\mathbf{R^{ij}}(\theta_{ij})$ ($i,j=1,\ldots,4$ and $i<j$) is the rotation matrix in the $ij$-plane over the angle $\theta_{ij}$. The rotation $\mathbf{R^{ij}_{\delta}} (\theta)$ contains CP-violating phase $\delta_{ij}$ in such a way that $\sin \theta_{ij} \rightarrow \sin \theta_{ij} e^{- i \delta_{ij}}$ and $- \sin \theta_{ij} \rightarrow - \sin \theta_{ij} e^{i \delta_{ij}}$. In Eq.~(\ref{eq:U}), $\delta_{13}$ is the usual CP-violating phase of the $3\nu$ scheme; whereas $\delta_{14}$ and $\delta_{24}$ are the two new phases. In this parametrization:
$$
U_{\mu 4}= \cos\theta_{14} \sin\theta_{24}e^{-i\delta_{24}} \quad , \quad U_{\tau 4} 
= \cos\theta_{14}\cos\theta_{24}\sin\theta_{34}~.
$$

The Hamiltonian describing propagation of this system in matter is 
\begin{equation}
\label{eq:hamiltonian}
\mathcal{H}= \frac{1}{2E_\nu}\mathbf{U_4} \mathbf{M^2}\mathbf{U_4}^\dagger+\mathbf{V}(r)~,
\end{equation}  
where $\mathbf{M^2}$ is the diagonal matrix of mass-squared differences:
\begin{equation}
\mathbf{M^2} \equiv  \mathbf{{\rm diag}}\left(0,\Delta m_{21}^2, \Delta m_{31}^2, \Delta m_{41}^2\right), 
~~~~\Delta m_{ij}^2\equiv m_i^2-m_j^2~,
\end{equation}
with $\Delta m_{41}^2\sim\mathcal{O}(1)~{\rm eV}^2$. The matrix of matter potentials in the flavor basis, $\mathbf{V}(r)$, after subtracting the neutral current contribution, takes the following form:
\begin{equation}
\mathbf{V}(r) = 
\sqrt{2}G_F\mathbf{ {\rm diag}}\left(N_e(r), 0,0, N_n(r)/2\right)~.
\end{equation}
Here $G_F$ is the Fermi constant; $N_e(r)$ and $N_n(r)$ are the electron and neutron number densities.

To avoid conflict with cosmological constraints we assume $\Delta m_{41}^2>0$, and therefore the resonance enhancement of oscillations takes place in the anti-neutrino channel, which leads to a dip in  $P(\bar{\nu}_\mu\to\bar{\nu}_\mu)$ at energies $\sim 4~{\rm TeV}~(\Delta m_{41}^2/{\rm eV}^2)$.

The appearance signals of LSND/MiniBooNE depend on $U_{\mu 4}$ and $U_{e 4}$ and do not depend on $U_{\tau 4}$. The disappearance depends on $U_{\mu 4}$ only. This is because the effects in these experiments are due to short baseline oscillations in vacuum. In contrast, the IceCube signal depends besides $U_{\mu 4}$ on the admixture $U_{\tau 4}$~\cite{Razzaque:2011ab,Esmaili:2012nz}. Therefore to get limit on the LSND/MiniBooNE sterile neutrino one needs to select the value of $U_{\tau 4}$ which leads to the smallest effect in IceCube for fixed $U_{\mu 4}$. The current upper limit on $U_{\tau 4}$ is rather weak: $\sin^2 2\theta_{34}<0.6$ at 90\% C.L.~\cite{Adamson:2011ku}. The global analysis of the oscillation data, which includes also the atmospheric neutrinos gives moderately stronger upper limit: $\sin^2 2\theta_{34}<0.6$ at $2\sigma$~\cite{Kopp:2013vaa}. IceCube itself can constrain $\sin^2 2\theta_{34}$ down to $\sim 0.1$ by using cascade events induced by the atmospheric neutrinos~\cite{Esmaili:2013cja}. Indeed, $\theta_{34} \neq 0$ leads to the oscillation transitions $\nu_\mu \rightarrow \nu_\tau$, $\nu_\tau$'s produce $\tau$ leptons and then decays of $\tau$'s generate cascades. Using the cascade events, it is also possible to disentangle the effects of $\theta_{24}$ and $\theta_{34}$. Indeed, $\theta_{24}\neq0$ results in a deficit of cascades, whereas $\theta_{34}\neq0$ via the $\nu_\mu \rightarrow \nu_\tau$ oscillations leads to an excess of cascades. In what follows we will explore the range $\sin^2 2\theta_{34} = 0 - 0.5$.

Let us consider the differences between the LSND/MiniBooNE and IceCube signals in more details. We will concentrate on the $\nu_\mu-$ charged current events. There are two simplifying circumstances at high energies: 

\begin{itemize}
\item
Strong suppression of the $\nu_e-$mixing everywhere apart from the $\nu_e - \nu_s$ resonance in the TeV range. Even in the resonance the  $\nu_e - \nu_\mu$ oscillations relevant for the $\nu_\mu-$events will be suppressed by small $U_{\mu 4}$. 

\item
Smallness of the original atmospheric $\nu_e-$flux at high energies.  
\end{itemize}

Under these circumstances one can exclude the $\nu_e-$flavor from consideration and neglect the 1-2 mass splitting. As a result,  the $4 \nu-$system is reduced to the system of three flavors $\nu_f = (\nu_\mu, \nu_\tau, \nu_s)$ mixed in three mass states $\nu_{\rm mass} = (\nu_2, \nu_3,  \nu_4)$: $\nu_f = {\bf U_3} \nu_{\rm mass}$. Here $\mathbf{U_3}$ is the $3\times3$ submatrix of $\mathbf{U_4}$ after removing the first row and column and setting $\theta_{1i}$ ($i=2,3,4$) to zero. In the next subsection we will make further simplification developing a single $\Delta m^2$ approximation. We will use these simplifications in our qualitative analysis which will allow to understand various results of numerical computations. The latter have been done for the complete $4\nu-$ system.  

\subsection{Single $\Delta m^2$ approximation}\label{sec:app}

At very high energies $E_\nu > (0.5 - 1)$~TeV, when $\Delta m_{31}^2$ also can be neglected, the system is described by $\Delta m_{41}^2$ and the vector of mixing parameters $\vec{U}^{T} = (U_{\mu 4}, U_{\tau 4},  U_{s 4})$. We can perform rotation $R_{\mu \tau} (\theta^{\prime})$ in the $\nu_\mu - \nu_{\tau}$ plane by the angle $\theta^{\prime}$,  determined by 
\be
\tan \theta^{\prime} = \frac{U_{\mu 4}}{U_{\tau 4}}~,
\label{eq:prime}
\ee
such that in new basis, $\nu_f^{\prime} = (\nu_{\mu}^{\prime}, \nu_\tau^{\prime}, \nu_s)$, the first component of the vector vanishes:
\be
\vec{U}^{\prime T} = \left(0, \sqrt{U_{\tau 4}^2 + U_{\mu 4}^2},  U_{s4}\right)~. 
\ee
In this basis the Hamiltonian becomes 
\be
H = \frac{\Delta m_{42}^2}{2E_\nu} \vec{U}^{\prime} \vec{U}^{\prime T} + V~. 
\label{hamilt}
\ee
The state $\nu_{\mu}^{\prime}$ decouples and the problem is reduced to two neutrinos problem with the mixing parameter in vacuum
\be
\sin^2 2 \theta_x = 4 U_{s4}^2 (U_{\tau 4}^2 + U_{\mu 4}^2) = 
4 U_{s4}^2 (1 - U_{s4}^2)~,  
\label{sinx}
\ee
and the potential $V$. 

Let us introduce the $S-$matrix in the basis $\nu_f^{\prime}$:  
\be
S^{\prime} = 
\left(\begin{array}{ccc}
1  &  0 & 0 \\
0 & A_{\tau^{\prime} \tau^{\prime}}  &  A_{s \tau^{\prime}} \\
0 &  A_{\tau^{\prime} s}             &  A_{ss}  
\end{array}
\right)~.
\label{eq:tilds1}
\ee
In terms of $S^{\prime}$ the $S$ matrix in the original flavor basis equals
\be 
S = R_{\mu \tau}(\theta^{\prime})^{\dagger} S^{\prime} R_{\mu \tau}(\theta^{\prime})~.  
\label{eq:Sflav}
\ee
From Eqs.~(\ref{eq:tilds1}), (\ref{eq:Sflav}) and (\ref{eq:prime}) we find the $\nu_\mu-$survival probability
\be
P_{\mu \mu} \equiv |S_{\mu \mu}|^2 =   
\left| 1 - \frac{U_{\mu 4}^2}{U_{\tau 4}^2 + U_{\mu 4}^2}
(1 - A_{\tau^{\prime} \tau^{\prime}} )\right|^2~. 
\label{pmumu}
\ee
It can be rewritten as 
\be
P_{\mu \mu} =
\left| 1 - \kappa (1 - A_{\tau^{\prime} \tau^{\prime}} )\right|^2~, 
\label{pmumu}
\ee
where the prefactor of the amplitude term, $(1 - A_{\tau^{\prime} \tau^{\prime}})$, equals 
\be
\kappa \equiv \frac{U_{\mu 4}^2}{U_{\tau 4}^2 + U_{\mu 4}^2} = \sin^2 \theta^{\prime}~.
\ee

Let us consider properties of the probability $P_{\mu \mu}$. Independently of the $U_{\tau 4}^2$, the maximal value $P_{\mu \mu}^{max} = 1$ is achieved if $A_{\tau^{\prime} \tau^{\prime}} = 1$. The minimal possible value (maximal oscillation effect) do depend on $U_{\tau 4}^2$: 
\begin{itemize}

\item
For $U_{\tau 4}^2 \leq U_{\mu 4}^2$, that is $\kappa \geq 1/2$, we have 
\be  
P_{\mu \mu}^{min} = 0~,  ~~~{\rm if}~~ A_{\tau^{\prime} \tau^{\prime}} = 
- \frac{U_{\tau 4}^2}{U_{\mu 4}^2}~.  
\label{small}
\ee  

\item
For $U_{\tau 4}^2 > U_{\mu 4}^2$  or  $\kappa < 1/2$, 
\be  
P_{\mu \mu}^{min} = |1 - 2 \kappa |^2 = 
\left|\frac{U_{\tau 4}^2 - U_{\mu 4}^2}{U_{\tau 4}^2 + U_{\mu 4}^2}\right|^2 , 
 ~~~{\rm if}~~~~ A_{\tau^{\prime} \tau^{\prime}} = -1~. 
\label{large}
\ee
\end{itemize}
Notice that the amplitude $A_{\tau^{\prime} \tau^{\prime}}$ depends on the neutrino energy, zenith angle, {\it etc.}, and conditions on the amplitude in the Eqs.~(\ref{small}) and (\ref{large}) may not be satisfied. So, extrema may not be realized and $P_{\mu \mu} > P_{\mu \mu}^{min}$.

For the $\nu_\mu - \nu_s$ mixing scheme, when $U_{\tau 4} = 0$, we obtain from Eq.~(\ref{pmumu})
\be
P_{\mu \mu} = \left| A_{\tau^{\prime} \tau^{\prime}} \right|^2~. 
\ee
If $U_{\tau 4} =  U_{\mu 4}$ (the ``$\nu_s$ - mass'' mixing scheme), 
\be
P_{\mu \mu} = \frac{1}{4}\left|1 + A_{\tau^{\prime} \tau^{\prime}} \right|^2~.
\ee

In the LSND and MiniBooNE experiments the oscillations occur in vacuum. For $V = 0$ the Hamiltonian in Eq.~(\ref{hamilt}) has the eigenvalues $\lambda_4 = \Delta m_{42}^2/2E_\nu$ and $\lambda_3 = 0$. Using Eq.~(\ref{hamilt}) it is straightforward to find that 
\be
A_{\tau^{\prime} \tau^{\prime}}^{vac} = 
1 - (U_{\tau 4}^2 + U_{\mu 4}^2) 
\left(1 - e^{-i\phi_4} \right)~, 
\label{Avac}
\ee
where the vacuum phase equals 
\be
\phi_4  \equiv \frac{\Delta m^2_{42} L}{2E_\nu}~.  
\ee
Inserting Eq.~(\ref{Avac}) into Eq.~(\ref{pmumu}) we obtain 
\be
P_{\mu \mu} =  \left| 1 - U_{\mu 4}^2
(1 - e^{-i\phi_4}) \right|^2~. 
\label{pmumuvac}
\ee
Here the dependence on $U_{\tau 4}^2$ disappears and Eq.~(\ref{pmumuvac}) coincides with the standard $2\nu$ oscillation probability which depends on $U_{\mu 4}^2$ only. 

In the case of IceCube, neutrinos oscillate in the matter of Earth and one should use the amplitude of oscillation in matter $A_{\tau^{\prime} \tau^{\prime}}^{vac} \rightarrow A_{\tau^{\prime} \tau^{\prime}}$. Cancellation of the factors which depend on $U_{\tau 4}^2$ does not occur anymore and the probability depends on $U_{\tau 4}^2$. Let us elucidate this dependence considering oscillations in matter with constant density. The amplitude $A_{\tau^{\prime} \tau^{\prime}}$ is function of $\sin^2 2 \theta_x$, $\Delta m_{42}^2$ and $V $. Both the oscillation depth and length depend on $U_{\tau 4}^2$ in non-trivial ways:  
\be
A_{\tau^{\prime} \tau^{\prime}} = 
\cos^2 \theta_x^m  e^{-i\phi_3^m} + \sin^2 \theta_x^m e^{-i\phi_4^m} 
= e^{-i\phi_3^m} \left[ \cos^2 \theta_x^m   + \sin^2 \theta_x^m e^{-i\Delta \phi^m}\right]~,
\label{Amatt0}
\ee
or
\be
1 - A_{\tau^{\prime} \tau^{\prime}} = 
1 - e^{-i\phi_3^m} + 
\sin^2 \theta_x^m \left( e^{-i\phi_3^m} - e^{-i\phi_4^m} \right)~.
\label{Amatt}
\ee
Here 
\be
\sin^2 \theta_x^m = \frac{1}{2} \left[ 
1 - \frac{1 + \tilde{V} -2 s^2_x}{
\sqrt{(1 + \tilde{V})^2 - 4 s^2_x \tilde{V}} 
}\right]~,
\label{sinxm}
\ee
and
\be
s^2_x \equiv \sin^2 \theta_x = (U_{\tau 4}^2 + U_{\mu 4}^2)~,
\label{sinx}
\ee
\be
\tilde{V} \equiv \frac{2E_\nu V}{\Delta m_{42}^2}~.
\ee
The phases equal
\be
\phi^m_{3,4} =
\frac{\phi_4}{2}
\left[1 + \tilde{V} \mp \sqrt{(1 + \tilde{V})^2 - 
4 s^2_x \tilde{V}}     
\right]~.
\label{phases}
\ee
(At the same time the prefactor $\kappa$ in Eq.~(\ref{pmumu}) is function of mixing parameters in vacuum.)

Let us analyze these expressions in two cases.  

1. Tails of resonances (energy range far from resonance) and $s_x^2 \ll 1$. In the lowest order, neglecting $4 \tilde{V} s^2_x$ under squared root in Eq.~(\ref{sinxm}), we have 
\be
\sin^2 \theta_x^m \approx 
\frac{s^2_x}{1 + \tilde{V}}~,
\label{sinxm1}
\ee 
\be
\phi_4^m = \phi_4 (1 + \tilde{V}),~~~~ \phi_3^m = 0~.
\ee
Then the probability equals 
\be
P_{\mu \mu} =  \left| 1 - \frac{U_{\mu 4}^2}{1 + \tilde{V}}
(1 - e^{-i\phi_4^m}) \right|^2~,
\label{pmumumat0}
\ee
which does not depend on $U_{\tau 4}^2$. 
  
Taking the first approximation in $s^2_x$ in the denominator of Eq.~(\ref{sinxm}) we obtain 
\be
\sin^2 \theta_x^m \approx 
\frac{s^2_x}{(1 + \tilde{V})^2 - 2s^2_x \tilde{V} }~,
\label{sinxm2}
\ee 
\be
\phi_3^m = \phi_4 \frac{\tilde{V}}{1 + \tilde{V}} s^2_x , ~~~~ 
\phi_4^m = \phi_4 (1 + \tilde{V}) - \phi_3^m~.
\ee

If $\phi_3^m \ll 1$, we have 
\be
\frac{1}{s^2_x}(1 - A_{\tau^{\prime} \tau^{\prime}}) = 
i \phi_4 \frac{\tilde{V}}{1 + \tilde{V}} + 
\frac{1 - e^{-i\phi_4 (1 + \tilde{V})}  - i \phi_4 \frac{\tilde{V}}{1 + \tilde{V}} 
s^2_x \left[1 +  e^{-i\phi_4 (1 + \tilde{V})}\right]} 
{(1 + \tilde{V})^2 - 2 \tilde{V}s^2_x}~. 
\label{Amatt2}
\ee
Dependence on $U_{\tau 4}^2$ appears only in corrections via $s^2_x$. In the neutrino (non-resonance) channel  we have the same expression with $\tilde{V} > 0$.

Notice that $|A_{\tau^{\prime} \tau^{\prime}}|^2$ is the standard $2\nu$ probability which depends on the phase difference $\phi_4^m - \phi_3^m$. In contrast, $P_{\mu \mu}$ is generically the probability of $3\nu$ system even in the limit $\Delta m_{31}^2 = 0$. As a result it depends on both phases $\phi_4^m$ and  $\phi_3^m$ separately. Nonzero value of $\phi_3^m$ leads, in particular, to the fact that $A_{\tau^{\prime} \tau^{\prime}}$ in Eq.~(\ref{Amatt0}) cannot be $1$ and to deviation of maximal value of the probability $P_{\mu \mu}$ from 1 (see Fig.~\ref{fig:prob-antinu}). From these formulae one can see that $P_{\mu \mu}$ decreases with the increase of $s_x^2$, and therefore,  $U_{\tau 4}$.

2. The resonance region. For $\tilde{V} = - 1$ we have  
\be
\sin^2 \theta_x^m \approx 
\frac{1}{2}(1 + s_x)~, 
\label{sinxm3}
\ee 
\be
\phi_{3,4}^m = \mp \phi_4 s_x~. 
\ee
(In resonance $\sin^2 \theta_x^m = \frac{1}{2}$.) Inserting these expressions into Eq.~(\ref{Amatt}) and Eq.~(\ref{pmumu}) we obtain
\be
P_{\mu \mu} =  \left| 1 - 
U_{\mu 4}^2 \frac{1}{s^2_x} \left[1 - \cos (\phi_4 s_x)
+ i s_x \sin (\phi_4 s_x)\right] \right|^2~. 
\label{pmumumat2}
\ee
With the increase of $U_{\tau 4}^2$, and consequently $s_x$, the probability in Eq.~(\ref{pmumumat2}) in resonance increases; {\it i.e.}, sterile neutrino effect decreases.

For small phases, $\phi_4 \sin \theta_x \ll 1$, Eq.~(\ref{pmumumat2}) leads to
\be
P_{\mu \mu} =  \left| 1 - 
U_{\mu 4}^2 (\phi_4^2 + i \phi_4) \right|^2~, 
\label{pmumumat3}
\ee
where again dependence on $U_{\tau 4}^2$ disappears. Notice that in the dip of $P_{\mu \mu}$ the phase equals $\phi_4 \sim \pi$, so for small ($\pi s_x$) the dependence of probability on the mixing scheme is very weak. It appears when $\sin \theta_x$ becomes large.

Summarizing, the difference between LSND/MiniBooNE and IceCube signals is due to matter effect. The dependence of probability on $U_{\tau 4}$ in IceCube signal can be understood in the following way. For fixed $U_{\mu 4}$ with the increase of $U_{\tau 4}$ the mixing angle $\theta_x$, which determines the strength of resonance (its width) and also the oscillation length, increases. However, for small $\sin \theta_x$ dependence of the oscillation probability on $U_{\tau 4}$ is weak. With increase of $U_{\tau 4}$ the mixing parameter $\sin \theta_x$ increases and dependence of probability on $U_{\tau 4}$ becomes strong: It suppresses the peak in the resonance region, since $|1 - A_{\tau^{\prime} \tau^{\prime}}|$ reaches maximal value and stops to increase, whereas the prefactor $\kappa$ continues to decrease. In contrast, beyond the resonance, in the tails, $|1 - A_{\tau^{\prime} \tau^{\prime}}|$ continue to increase due to widening of the resonance. As a result, here the disappearance probability increases with $U_{\tau 4}$ and the minimal effect is for $U_{\tau 4}= 0$ (flavor mixing scheme). This approximate analytical consideration helps to understand various features of exact numerical results. 

\subsection{Survival probabilities}

We find the $\nu_\mu$ ($\bar{\nu}_\mu$) survival probabilities, as functions of neutrino energy and zenith angle, by solving the evolution equation with the Hamiltonian in Eq.~(\ref{eq:hamiltonian}) numerically. We take the best-fit values of active-active mixing angles as well as $\Delta m_{21}^2=7.4\times 10^{-5}~{\rm eV}^2$ and $\Delta m_{31}^2 = 2.4\times 10^{-3}~{\rm eV}^2$ according to~\cite{GonzalezGarcia:2012sz}. We use as a benchmark value $\sin^2 2\theta_{24} = 0.04$ or $U_{\mu 4}^2 = 0.01$ which is at the border of IceCube sensitivity region (see Sec.~IV). This value is substantially lower than the one required by LSND: $\sin^2 2\theta_{24} > 0.1$. 

\begin{figure}[t!]
\centering 
\includegraphics[width=0.7\textwidth]{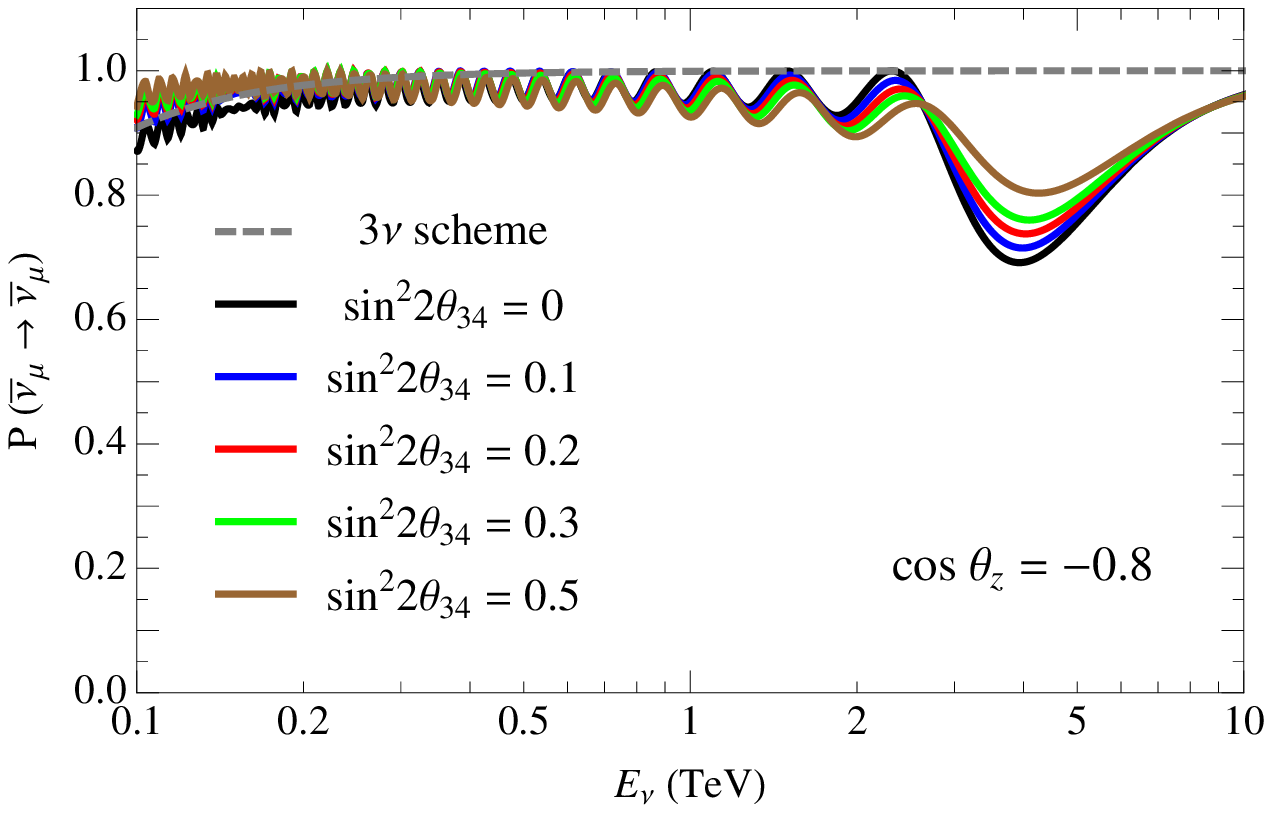}
\label{fig:pmubar,2}
\quad
\includegraphics[width=0.7\textwidth]{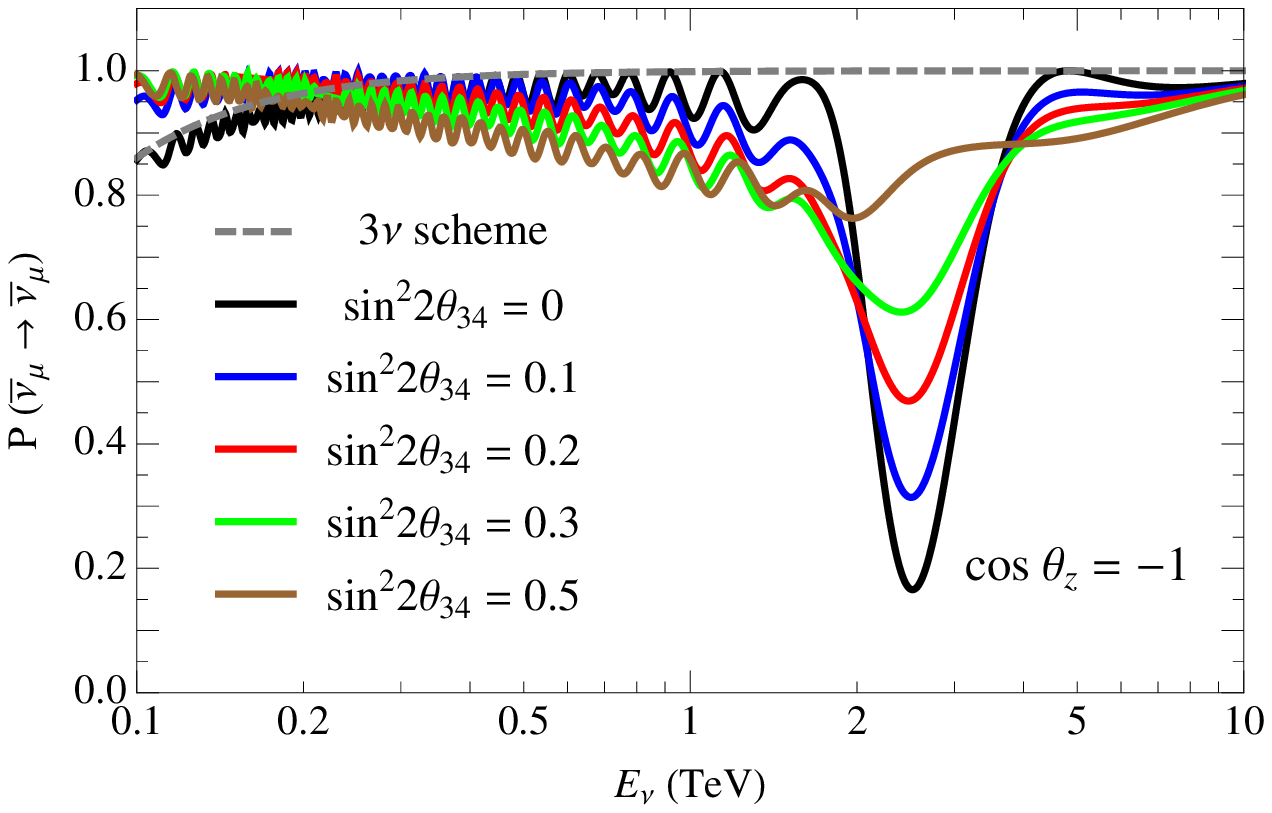}
\label{fig:pmubar,1}
\caption{\label{fig:prob-antinu}
The dependence of the $\bar{\nu}_\mu-$ survival probability on the neutrino energy for different values of $U_{\tau 4}$. We take $\Delta m_{41}^2=1~{\rm eV}^2$ and $\sin^2 2 \theta_{24} = 0.04$. Top panel is for $\cos \theta_z = - 0.8$ and bottom panel is for $\cos \theta_z = - 1$~.}
\end{figure}

\begin{figure}[t!]
\centering
\includegraphics[width=0.7\textwidth]{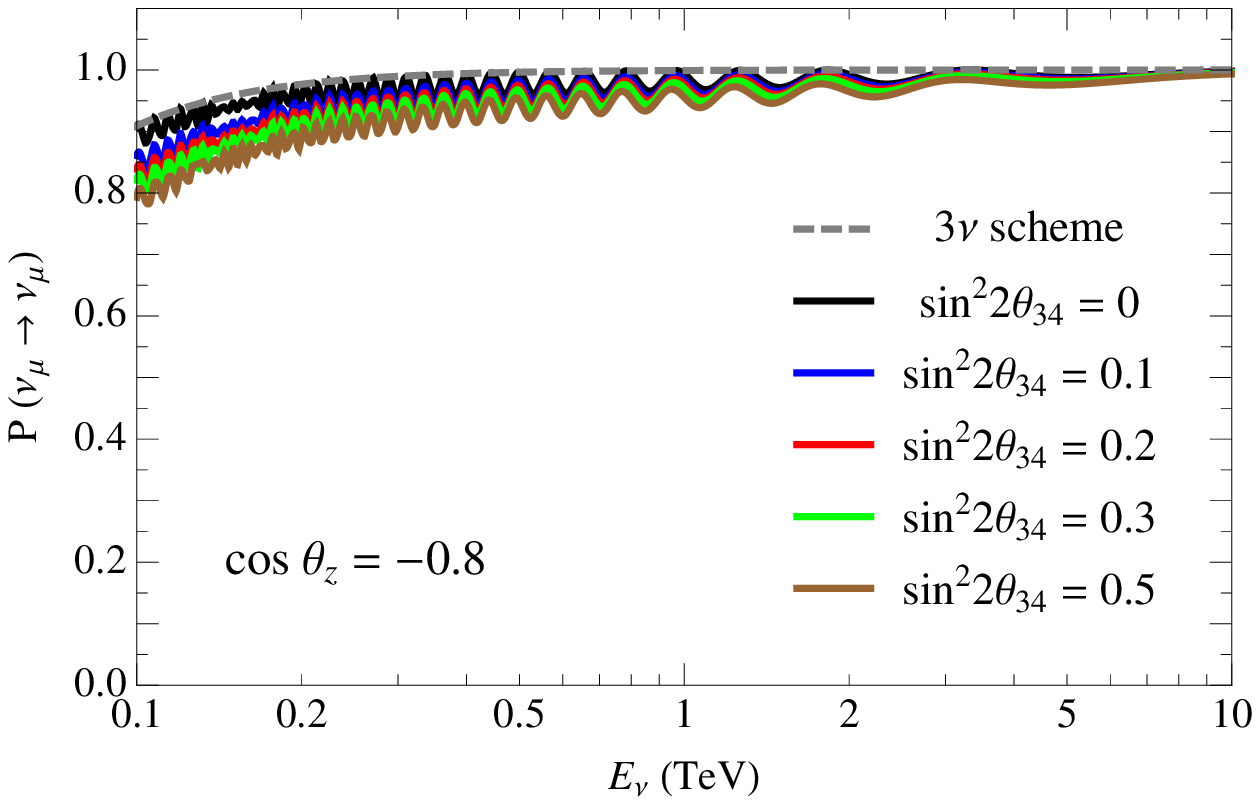}
\label{fig:pmu,2}
\quad
\includegraphics[width=0.7\textwidth]{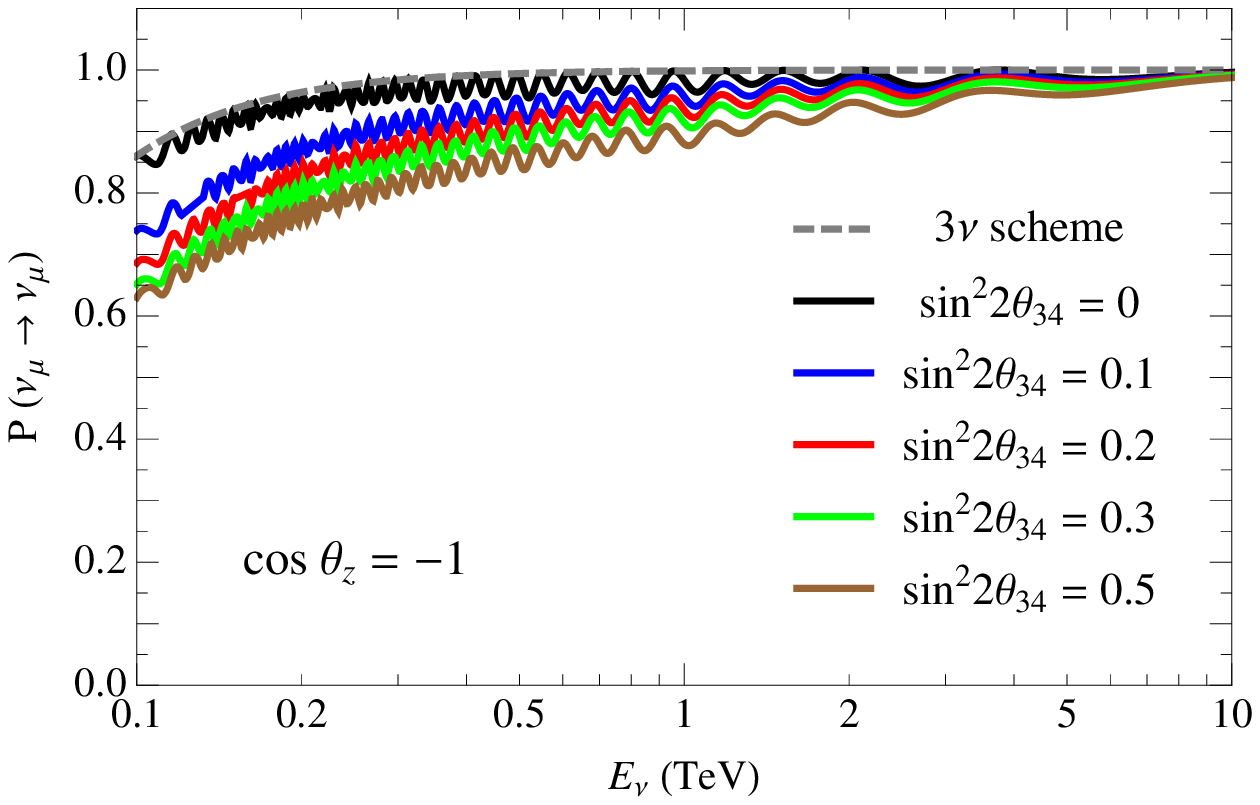}
\label{fig:pmu,1}
\caption{\label{fig:prob-nu}
The same as in Fig.~\ref{fig:prob-antinu} for the  $\nu_\mu-$ survival probability.}
\end{figure}

In Fig.~\ref{fig:prob-antinu} we show dependence of the $\bar{\nu}_\mu$ survival probability (resonance channel) on the neutrino energy for different values of $U_{\tau 4}^2$ and for two different zenith angles. In the case of $\cos \theta_{z} = -0.8$ (mantle crossing trajectory, top panel) the resonance dip appears at $E_\nu \approx 4$~TeV~(for $\Delta m^2_{41} = 1$~eV$^2$) as a result of resonance enhancement of oscillations. For $E_\nu > 0.4$~TeV the dependence is well described by the single $\Delta m^2$ approximation. Below $E_\nu= 0.4$~TeV effect of oscillations driven by $\Delta m^2_{31}$ becomes important. It increases with the decrease of energy as a result of interference of the $\Delta m^2_{31}$ and $\Delta m^2_{41}$ modes of oscillations. The dashed line in Fig.~\ref{fig:prob-antinu} shows the standard $3\nu-$ probability. Maximal effect of sterile neutrinos is in the resonance. Notice that the line describing probability in the presence of sterile neutrino for $U_{\tau 4}^2 = 0$ touches the $3\nu$ probability (the latter is the upper bound for the probability with sterile neutrino). With the increase of $U_{\tau 4}^2$ the sterile neutrino effect decreases in the resonance dip. It increases in the tail above $E_\nu \sim 0.4$~TeV (this can be seen from the analytical formulas obtained in Sec.~\ref{sec:app}), and it decreases again below $E_\nu \sim 0.4$~TeV -- in the region were standard oscillations become important. Also, with the increase of $U_{\tau 4}^2$ the oscillatory curve shifts slightly to higher energies. 

For $\cos \theta_{z} = -1$ (core crossing trajectory, bottom panel) the parametric enhancement of oscillations takes place. The parametric dip at $2.3$~TeV is larger than the resonance dip. The dependence of probability on $U_{\tau 4}^2$ is similar to that for the mantle crossing trajectories.

In Fig.~\ref{fig:prob-nu} we show the $\nu_\mu-$survival probabilities (non-resonance channel). In this channel oscillations are matter suppressed. Again below $0.4$~TeV the oscillation effect is driven by $\Delta m^2_{31}$. Mixing with sterile neutrinos enhances this effect, especially for large $U_{\tau 4}^2$. With increase of $U_{\tau 4}^2$ the probability decreases for all energies, so that the weakest $\nu_s$ effect is for $U_{\tau 4}^2 = 0$. 

Summarizing, the effect of sterile neutrino increases with the increase of $U_{\tau 4}^2$ at all energies apart from the region of resonance and parametric dips in the antineutrino channel, and as we will see, the former dominates in the integral effect. So, to exclude sterile neutrino mixing with IceCube one should consider the case $U_{\tau 4}^2= 0$.   

The highest acceptance of the IceCube detector for $\nu_\mu$ events is in the range (0.3 - 1)~TeV. Thus, for $\Delta m^2_{41} = 1$~eV$^2$ the maximum sensitivity is in the region of low energy tail of the survival probability, where the sterile neutrino effect increases with $U_{\tau 4}$. For smaller $\Delta m^2_{41}$ the resonance dip shifts to the region of the maximal IceCube acceptance and dependence of effect on $U_{\tau 4}$ can be opposite at least in the restricted energy range. However this opposite dependence disappears after smearing over the neutrino energies.

\subsection{CP-violation effects}\label{sec:cp}

Let us consider dependence of the sterile neutrino effects on the CP-phases. In $3\nu$ approximation of the parametrization of Eq.~(\ref{eq:U}) the mixing depends on one CP-phase $\delta_{24}$: 
\be
\label{eq:cp}
U_f = R^{34} R^{24}_{\delta} (\delta_{24}) R^{23}. 
\ee
Clearly, in the cases where mixing (in our approximation) can be parametrized by two rotations, there is no CP-violation, and no dependence on the effects of CP-phase. (Beyond this approximation the CP-violation will show up, however, its effects will be suppressed.)

In the case of $U_{\tau 4}=0$ the mixing matrix in Eq.~(\ref{eq:cp}) becomes $U_f  =  R^{24}_{\delta} R^{23}$. The CP-phase $\delta_{24}$ (as well as any other phase introduced here) can be eliminated by redefining the neutrino mass ($\nu_4 \rightarrow \nu_4 e^{i\delta_{24}}$), or/and  flavor fields. Therefore the CP-violation (in $3\nu$ approximation) should be proportional to $U_{\tau 4}$ or $\sin \theta_{34}$.

Furthermore, if $\Delta m^2_{31} = 0$, the 2-3 rotation can be removed, so that the mixing matrix in Eq.~(\ref{eq:cp}) is reduced to two rotations: $U_f  = R^{34} R^{24}_{\delta}$, and again the CP-phase becomes unphysical. The CP-violation effect should be proportional to the phase $\phi_{31}$ induced by $\Delta m^2_{31}$ or more precisely to $\sin \phi_{31}$ and to $\sin 2 \theta_{34}$. The CP-violation in the probability appears as an interference of the main term induced by $\Delta m^2_{41}$, which is close to 1 beyond the resonance, and the term induced by $\Delta m^2_{31}$. Therefore the CP-violating part of the survival probability should be approximately equal to
\bea
P_{\mu \mu}^{\delta} & \sim &  2 \cos \delta_{24} \sin 2 \theta_{34} \sin 2 \theta_{24}  \sin \phi_{31}
\approx 2 \cos \delta_{24} \sin 2 \theta_{34} \sin 2 \theta_{24}  \phi_{31}\nonumber \\ 
& = &  
\cos \delta_{24} \left(\frac{\Delta m^2_{41} L}{E_\nu}\right) \sin 2 \theta_{24} \sin 2 \theta_{34}~. 
\label{pmumud}
\eea
(It also is proportional to $\sin 2 \theta_{23} \approx 1$ which is omitted above.) Here we have taken into account also that the survival probability is a CP-even function of $\delta_{24}$. According to Eq.~(\ref{pmumud}) the CP-violation effect increases linearly with $\sin 2 \theta_{34}$ and also increases with decrease of energy. For high energies, where the 1-3 oscillation phase is very small, the effects are suppressed. These dependences can be seen in Fig.~\ref{fig:prob-cp}. In the top (bottom) panel of Fig.~\ref{fig:prob-cp} we show the dependence of $\nu_\mu$ ($\bar{\nu}_\mu$) survival oscillation probability on neutrino energy for various values of CP-phase $\delta_{24}$. The CP-violation effect appears below $0.7$~TeV where the $\Delta m_{31}^2$ driven oscillations become important. Notice that for $\delta_{24} = \pi/2$ the $\nu_\mu$ and $\bar{\nu}_\mu$ survival probabilities at low energies are nearly the same. For $\delta_{24} = \pi$ the $\nu_\mu-$  and $\bar{\nu}_\mu-$ probabilities switch with each other in comparison with $\delta_{24} = 0$ case: 
$$
P(\nu_\mu \rightarrow \nu_\mu;\delta = 0) \approx  
P(\bar{\nu}_\mu \rightarrow \bar{\nu}_\mu;\delta = \pi)~. 
$$
Varying $\delta_{24}$ from $0$ to $\pi$ we obtain from Eq.~(\ref{pmumud})  
$$
\Delta P_{\mu \mu}^{\delta} = 2 \left(\frac{\Delta m^2_{41} L}{E_\nu}\right) \sin 2 \theta_{24} \sin 2 \theta_{34}~. 
$$
For $E_\nu = 0.5$ TeV it gives $\Delta P_{\mu \mu}^{\delta} \sim 0.05$ in agreement with Fig.~\ref{fig:prob-cp}. 

Nonzero $\delta_{24}$ have opposite effects in neutrino and antineutrino channels: it enhances the oscillations due to $\Delta m_{31}^2$ in anti-neutrino channel and suppresses oscillations in neutrino channel. This leads to partial cancellation of the CP-phase effect at IceCube where signals from neutrino and antineutrino sum up. As a result (see Sec.~\ref{sec:dis}), CP-violation effect is subleading with respect to the effect of 3-4 mixing and our conclusion that the weakest limit on $U_{\mu 4}$ is realized for $U_{\tau4}=0$ still holds. 

\begin{figure}[t!]
\centering
\includegraphics[width=0.7\textwidth]{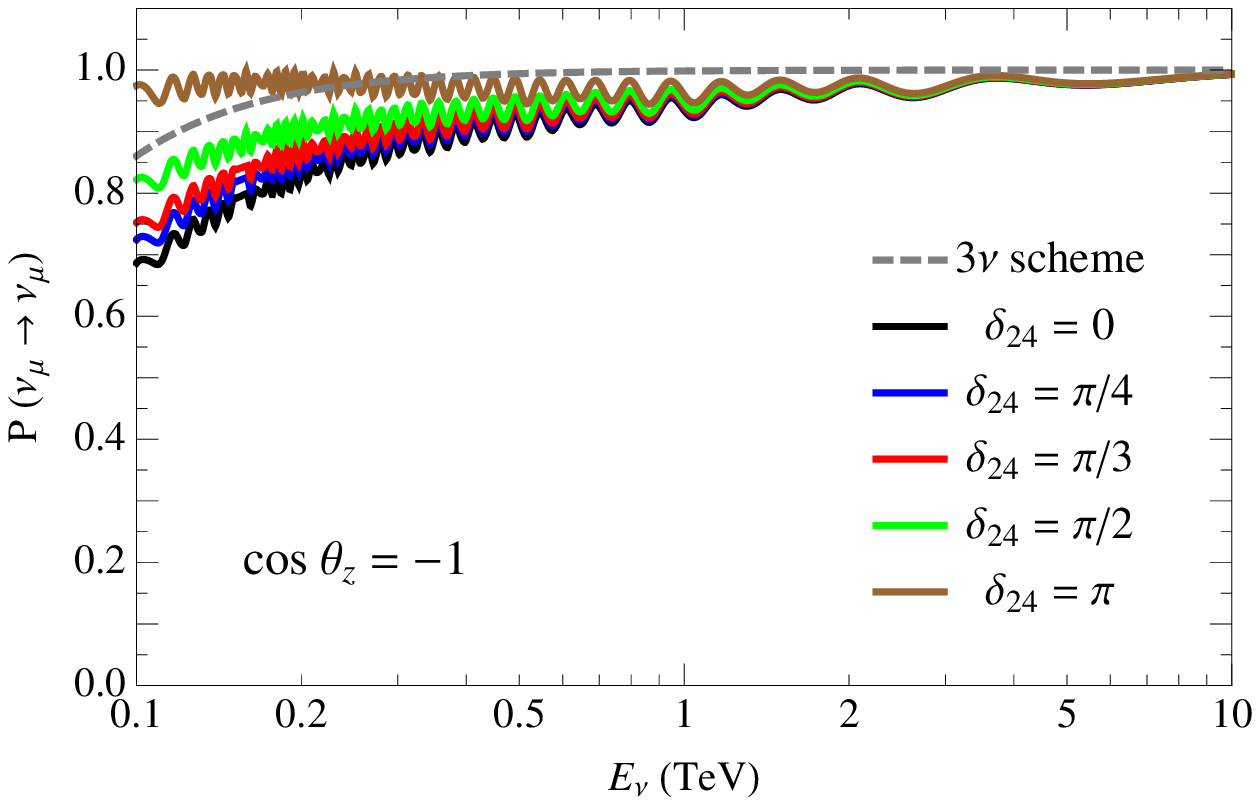}
\label{fig:pcp,1}
\quad
\includegraphics[width=0.7\textwidth]{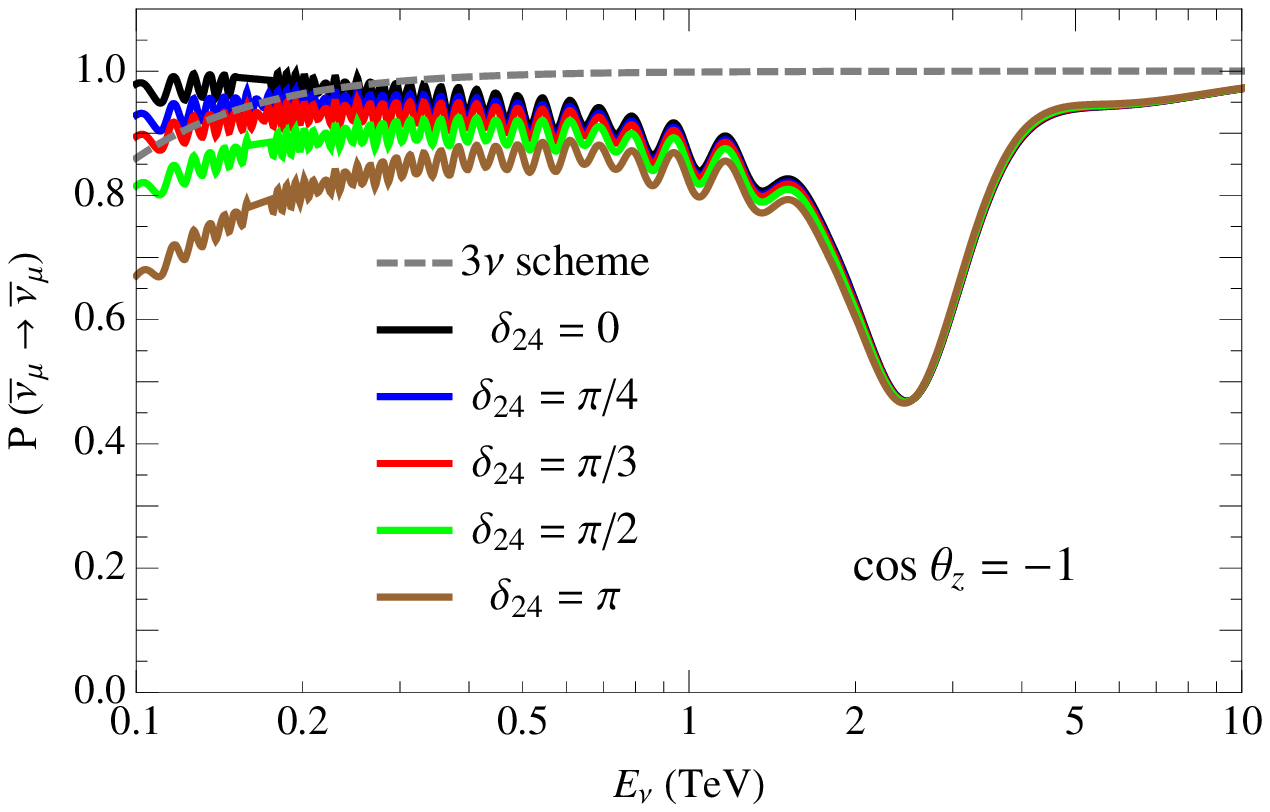}
\label{fig:pcp,2}
\caption{\label{fig:prob-cp}
The dependence of the $\nu_\mu$ (top panel) and $\bar{\nu}_\mu$ (bottom panel) survival probabilities on the neutrino energy for different values of CP-violating phase $\delta_{24}$ in Eq.~(\ref{eq:cp}). We take $\Delta m_{41}^2=1~{\rm eV}^2$, $\sin^2 2 \theta_{24} = 0.04$, $\sin^2 2 \theta_{34} = 0.2$ and $\cos \theta_z = - 1$.}
\end{figure}

\section{Zenith angle and energy distributions of the $\nu_\mu$ events}
\label{sec:dis}

The number of $\mu$-track events in IceCube originating from the $\nu_\mu$ ($\bar{\nu}_\mu$) CC interactions with the reconstructed neutrino energy and direction in the ranges $[E_\nu^r,E_\nu^r+\Delta_j E_\nu^r]$ and $[\cos\theta_z,\cos\theta_z+\Delta_i \cos\theta_z]$ is given by
\begin{eqnarray}
\label{eq:events,smear}
 N_{i,j} = T\Delta\Omega \int_{\Delta_i \cos\theta_z} {\rm d}\cos\theta_z \int_{\Delta_j E_\nu^r}  {\rm d}E_\nu^r \int {\rm d}E_\nu & \nonumber \\
 & \!\!\!\!\!\!\!\!\!\!\!\!\!\!\!\!\!\!\!\!\!\!\!\!\!\!\!\!\!\!\!\!\!\!\!\!\!\!\!\!\!\!\!\!\!\!\!\!\!\!\!\!\!\!\!\!\!\!\!\!\!\!\!\!\!\!\!\!\!\!\!\!\!\!\!\!\!\!\!\!\!\!\!\!\!\!\!\! G(E_\nu^r,E_\nu) A_{\rm eff}^{\nu_\mu} (E_\nu,\cos\theta_z) \left[ \sum_{\alpha=e,\mu}\Phi_{\nu_\alpha}(E_\nu,\cos\theta_z) P_{\alpha\mu} \right] + (\nu \to \bar{\nu} ) ,~
\end{eqnarray} 
where $A_{\rm eff}^{\nu_\mu(\bar{\nu}_\mu)}$ is the muon (anti)neutrino effective area and $\Phi_{\nu_\alpha}$ is the flux of atmospheric $\nu_\alpha$~\cite{Honda:2006qj,Athar:2012it}. As we mentioned above, in the energy range we are considering the contribution of $\Phi_{\nu_e}$ can be neglected since it is double suppressed: first, by small original $\Phi_{\nu_e}$ flux, and second, by small $\nu_e \rightarrow \nu_\mu$ oscillation probability which is proportional to $|U_{\mu 4}|^2$. In Eq.~(\ref{eq:events,smear}) $T$ is the exposure period and $\Delta\Omega=2\pi$ is the azimuthal acceptance of the IceCube detector. For estimations we use the IceCube-79 effective area in the energy range of $(0.1-10)$~TeV obtained by rescaling of the IceCube-40 effective area~\cite{Esmaili:2012nz,Abbasi:2010ie}. In our estimation we assume the same ratio of $A_{\rm eff}^{\bar{\nu}_\mu}/A_{\rm eff}^{\nu_\mu}$ as for the IceCube-40~\cite{Esmaili:2012nz}. We take 3 times larger statistics than IceCube-79; that is $T=3\times T_{\rm IC79}$ in Eq.~(\ref{eq:events,smear}), where $T_{\rm IC79}=318$~days. 

In Eq.~(\ref{eq:events,smear}), $E_\nu^r$ and $E_\nu$ are the reconstructed and true neutrino energies, and $G(E_\nu^r,E_\nu)$ is the resolution (reconstruction) function. The observable quantities in the IceCube detector are the energy and direction of muons produced in the CC interaction of $\nu_\mu$ and $\bar{\nu}_\mu$ with nuclei. Therefore, there are two contributions to the width of the reconstruction functions  of energy and direction of neutrinos: 1) the finite resolutions in measurement of the muon energy and direction; 2) the kinematic uncertainty related to the difference between muon energy and direction and the neutrino ones. Let us consider these contributions in order. In the TeV range the IceCube detector measures the muon direction with precision of less than $1^\circ$~\cite{Abbasi:2010ie}. The average angle between the neutrino and muon momenta in the CC interactions is $\theta\sim\sqrt{m_p/E_\nu}$ which decreases from $\sim5^\circ$ at $E_\nu=100$~GeV down to $\sim 0.5^\circ$ at $E_\nu=10$~TeV. Thus, we can identify the measured zenith angle of muon with the zenith angle of neutrino and consider 20 bins in $\cos\theta_z$ without any smearing of the distributions before binning.      

The reconstruction of neutrino energy is by far less precise. The observable quantity in the IceCube detector is the energy loss of muons, $dE_\mu/dx$, which is related to the muon energy $E_\mu$ by  
$$
\frac{dE_\mu}{dx}=-\alpha-\beta E_\mu~.  
$$
Here $\alpha$ and $\beta$ are nearly energy-independent coefficients describing the energy losses due to ionization and radiation respectively. For energies $\gtrsim 1$~TeV the radiation ($\beta$-term) dominates and the energy loss of muons is proportional to energy. Therefore the muon energy can be determined by measuring the energy loss even over a part of the muon track provided that the point (vertex) of muon production is known. However, in the high energy range for most of events the vertex of neutrino-nucleon interaction is outside of the geometrical volume of the detector which severely restricts the energy reconstruction and only lower bound can be established on the energy. For the low energy range, where the ionization ($\alpha$-term) dominates, the energy loss of muons is independent of their energy and the muon energy can be inferred from the energy loss measurements only when the whole track of muon is inside the geometrical volume of detector. At low energies most of the muons are produced inside the detector and therefore the muon energy reconstruction improves. The energy of neutrino is related to the muon energy through the inelasticity $y$ (fraction of the neutrino energy transferred to hadrons): $E_{\nu_\mu} = E_\mu/(1 - y)$. The average inelasticity $y \equiv E_{hadron}/E_{\nu_\mu}$, is nearly constant in the range $(0.1-10)$~TeV, however, $d\sigma^{\rm CC}/dy$ has wide $y-$distribution. When the vertex of $\nu_\mu$ CC interaction is inside the detector, measurement of the hadronic cascade energy would improve the neutrino energy reconstruction. Putting all these factors together, the IceCube collaboration claimed the resolution of the neutrino energy reconstruction $0.3$ in units of $\log_{10}(E_\nu/{\rm GeV})$ in the $(0.1-10)$~TeV range (see~\cite{Abbasi:2010ie}).

We compute the zenith angle distributions of events with and without sterile neutrinos using Eq.~(\ref{eq:events,smear}). We take as the reconstruction function $G(E_\nu^r,E_\nu)$ the normalized Gaussian distribution with width $\sigma_E=E_\nu$. The estimations show that variations of  $\sigma_E$ within $20\%$ do not produce significant changes of the  IceCube sensitivity. The reason is that although energy smearing decreases the depth of resonance dip in $P_{\mu\mu}$, it also widen the dip which can partially compensate the former. Since the acceptance of IceCube detector changes with energy, the smearing of distributions leads to a moderate weakening of bounds.  After the smearing of events, we integrated the number of events over the  energy bins $E_\nu^r$, as described below.

The zenith angle dependences of events are determined by the probabilities discussed in the previous section and the product $A_{\rm eff} \Phi_{\nu_\mu}$. The function $A_{\rm eff} \Phi_{\nu_\mu} (E_\nu)$ has maximum at $E_\nu \sim 0.2$~TeV. It decreases by one order of magnitude at $E_\nu = 2$~TeV and by another factor of 5 down to $E_\nu = 5$~TeV. On the other hand, features in the oscillation probabilities have $\log E_\nu$ scale. So, to take into account their contributions to the integral effect, the correct factor would be $E_\nu A_{\rm eff} \Phi_{\nu_\mu}$, which has maximum in the range $(0.5 - 1)$~TeV. In the resonance this product is only 2 times smaller than the maximum of $E_\nu A_{\rm eff} \Phi_{\nu_\mu}$.    

\begin{figure}[t!]
\centering
 \includegraphics[width=0.6\textwidth]{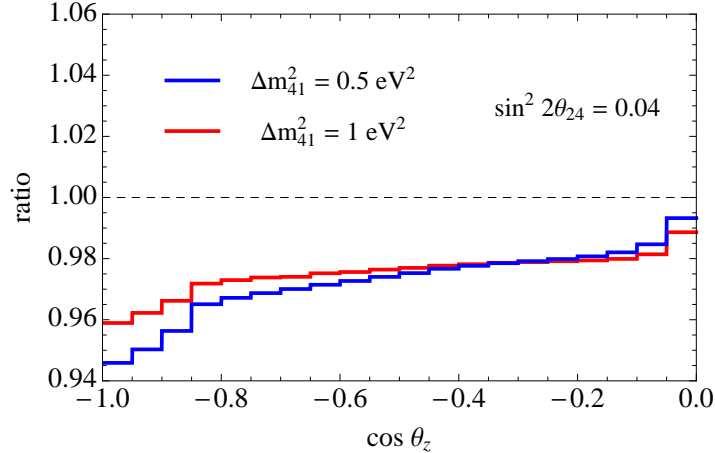}
\caption{\label{fig:dis,enuint}Distortion of the zenith angle distributions of $\mu$-track events due to the $\nu_\mu - \nu_s$ oscillations integrated over the energy interval $(0.1-10)$~TeV. Shown is the ratio of number of events with and without mixing with sterile neutrinos. We set $\sin^2 2\theta_{24}=0.04$ and the blue (red) curve corresponds to $\Delta m_{41}^2=0.5~{\rm eV}^2$ ($1~{\rm eV}^2$).}
\end{figure}

In Fig.~\ref{fig:dis,enuint} we show distortion of the zenith angle distribution of the $\mu$-track events, that is, the ratio of the distributions with and without sterile neutrinos. We take the benchmark value  $\sin^2 2\theta_{24} = 0.04$ and $U_{\tau4} = 0$. The events are integrated over whole energy range $(0.1-10)$~TeV. Fig.~\ref{fig:dis,enu2bin} shows distortion of the distributions in two energy bins: $[0.1,1]$~TeV and $[1,10]$~TeV. Fig.~\ref{fig:dis,enu3bin} is for three energy bins: $[0.1,0.4]$~TeV, $[0.4,1.8]$~TeV and $[1.8,10]$~TeV. The numbers of events in these regions are 9532, 14277 and 9139 correspondingly for the IceCube-79 exposure. This gives an idea about the relative contributions of these three regions.

With the decrease of $\Delta m^2_{41}$ the resonance and parametric dips shift to lower energies towards the maximum of function $E_\nu A_{\rm eff} \Phi_{\nu_\mu}$. This leads to stronger distortion of the distributions. Notice that for $\Delta m^2_{41} = 0.5$~eV$^2$ the center of the parametric dip is at $E_\nu = 1.2$~TeV. Therefore in the case of three bins (see Fig.~\ref{fig:dis,enu3bin}) the center is in the second bin, whereas for $\Delta m^2_{41} = 1.0$~eV$^2$ it is in the third bin. As a result, the effect for core crossing trajectories is stronger in the second bin for $\Delta m^2_{41} = 0.5$~eV$^2$ and in the third bin for $\Delta m^2_{41} = 1.0$~eV$^2$.

\begin{figure}[t]
\centering
 \includegraphics[width=0.5\textwidth]{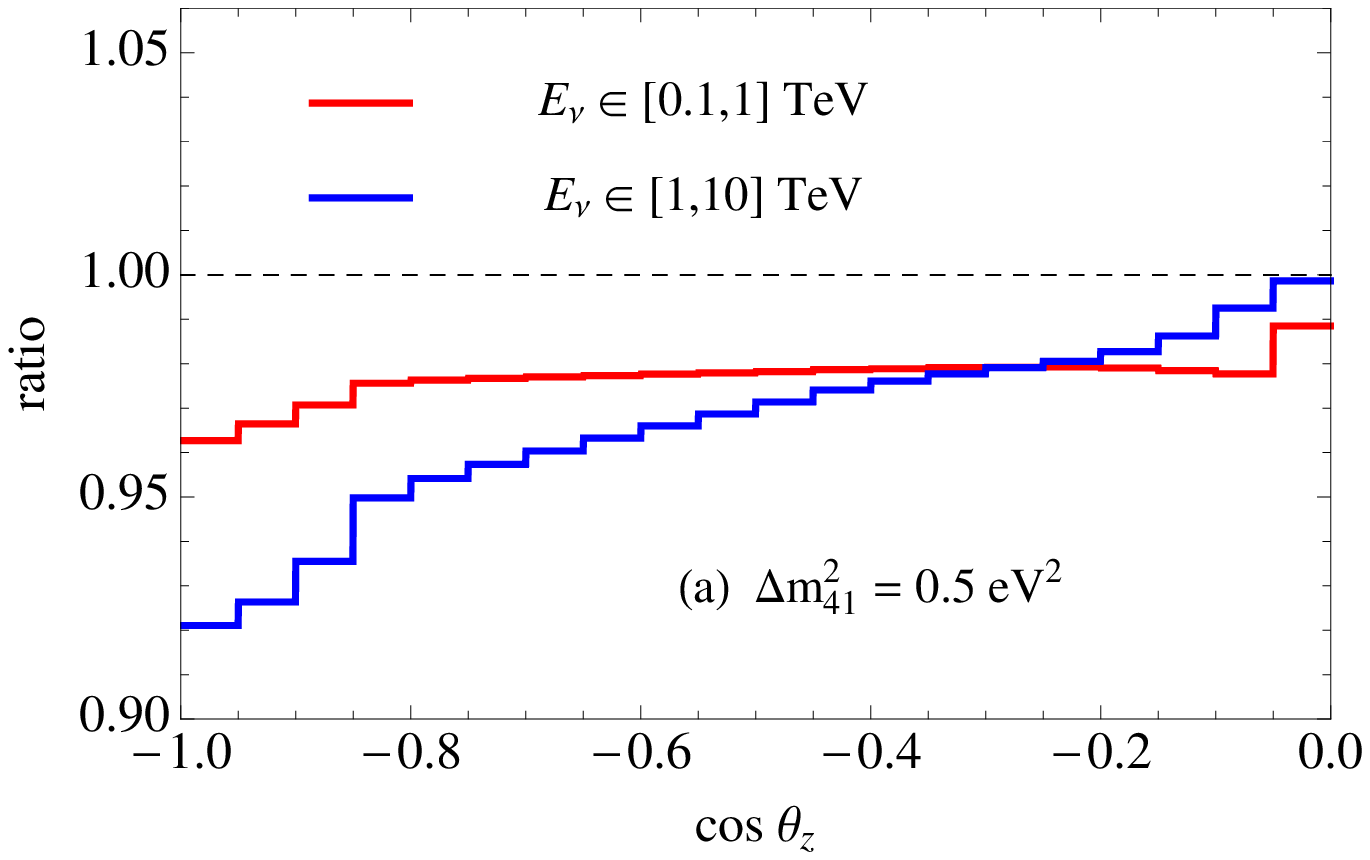}\includegraphics[width=0.5\textwidth]{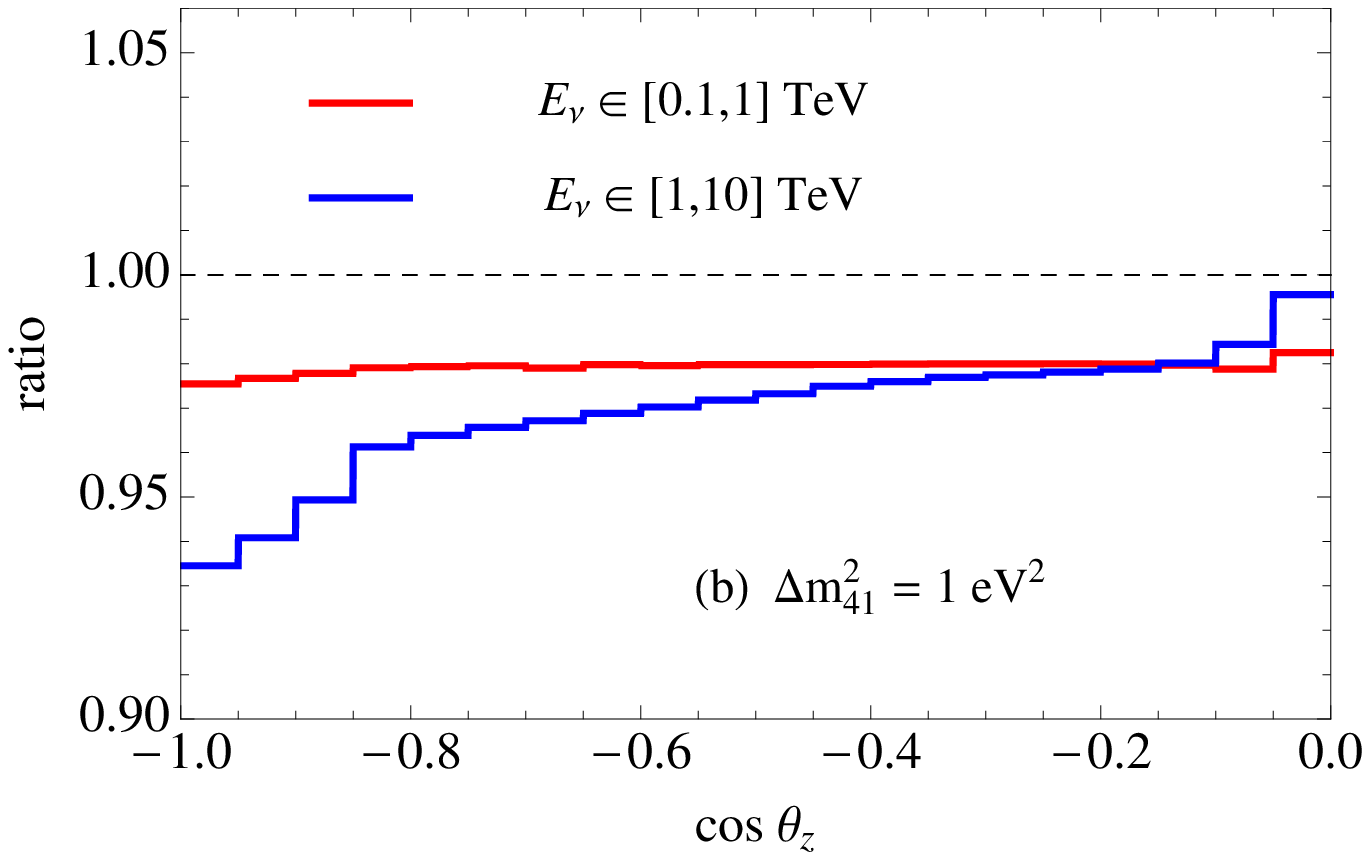}
\caption{\label{fig:dis,enu2bin}The same as Fig.~\ref{fig:dis,enuint}, with numbers of events integrated over the energy bins $(0.1-1)$~TeV and $(1-10)$~TeV. a) $\Delta m^2_{41} = 0.5$~eV$^2$, and b) $\Delta m^2_{41} = 1$~eV$^2$.}
\end{figure}
\begin{figure}[t!]
\centering
 \includegraphics[width=0.5\textwidth]{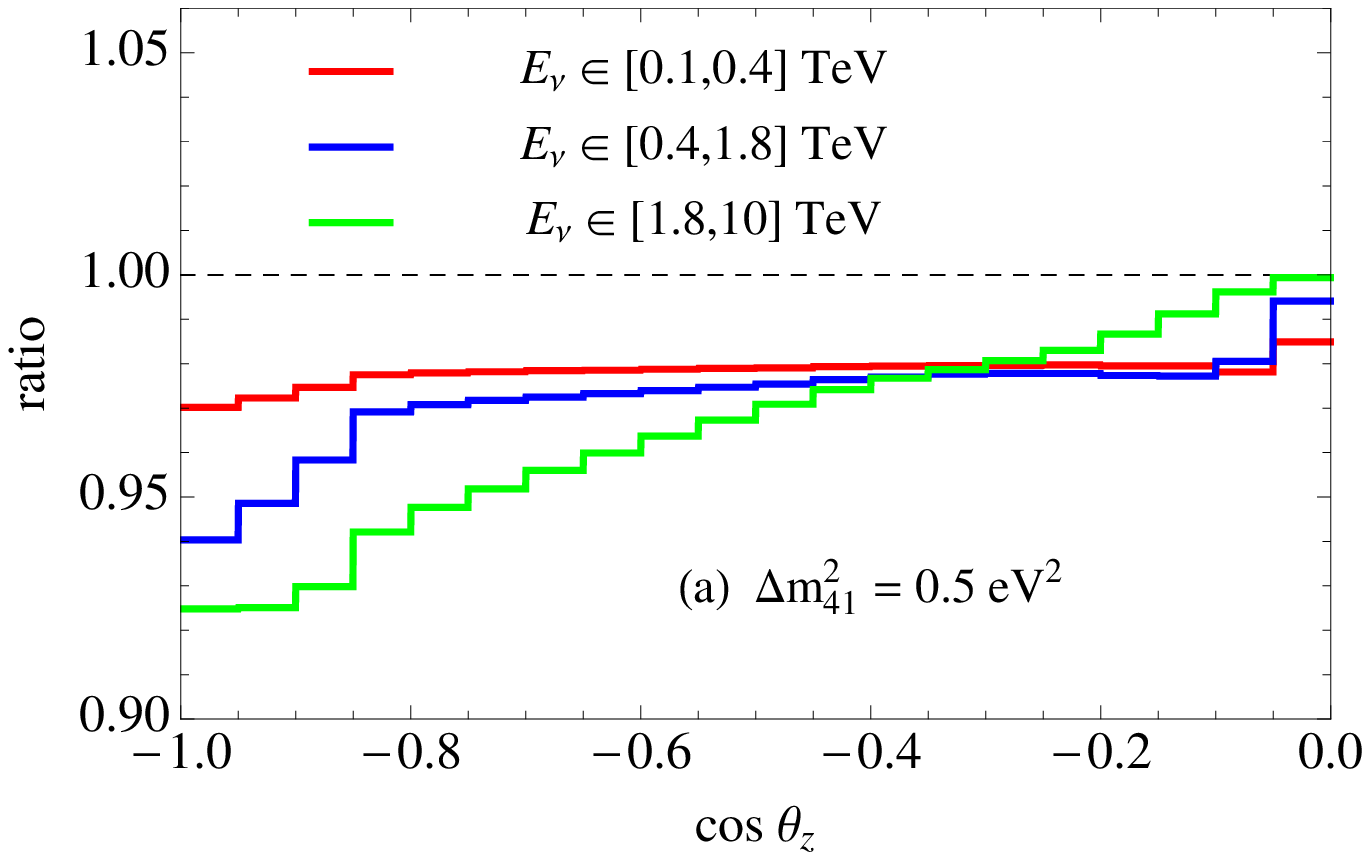}\includegraphics[width=0.5\textwidth]{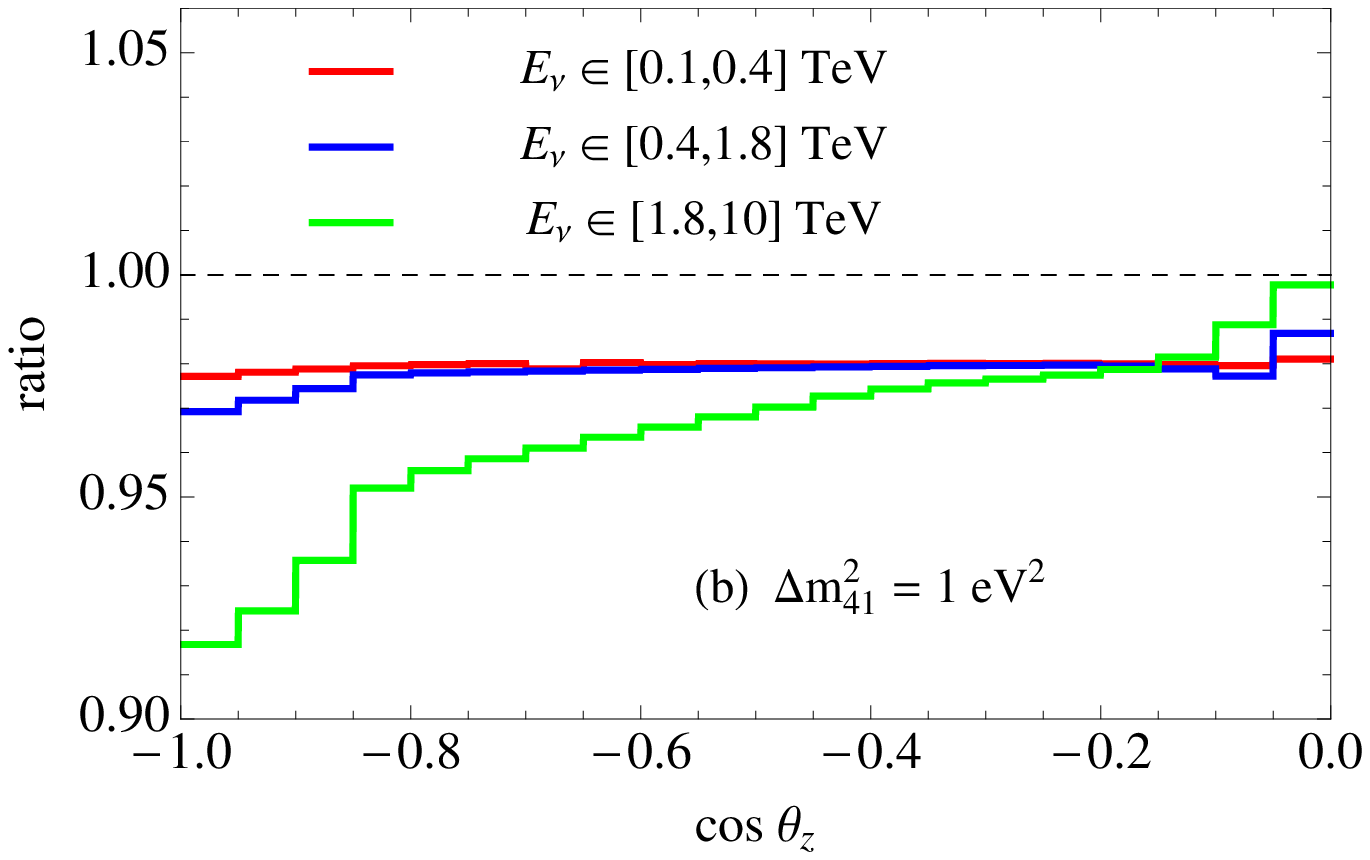}
\caption{\label{fig:dis,enu3bin}The same as Fig.~\ref{fig:dis,enuint}, with integration over the neutrino energy bins: $(0.1-0.4)$~TeV, $(0.4-1.8)$~TeV and $(1.8-10)$~TeV. a) $\Delta m^2_{41} = 0.5$~eV$^2$, and b) $\Delta m^2_{41} = 1$~eV$^2$.}
\end{figure}

The suppression of the number of muon-track events increases with the decrease of $\cos \theta_z$; {\it i.e.}, with approaching to vertical trajectories. A jump in the distributions at $-\cos \theta_z = (0.05 - 0.10)$ is related to turning on the oscillation effect at low energies. The position of the first oscillation minimum in the survival probability is determined by the condition $\cos \theta_z \approx - l_\nu/2 R_{E} $, where $R_E $ is the radius of the Earth and $l_\nu = 4\pi E_\nu / \Delta m^2_{41}$ is the vacuum oscillation length (at low energies the matter effect is small). With further decrease of $\cos \theta_z$ (increase of the length of neutrino trajectory) the oscillations become quickly averaged. For $E_\nu = 0.2$ TeV and $\Delta m^2_{41} = 0.5$ ${\rm eV}^2$ the condition gives $\cos \theta_z = - 0.04$ for the position of minimum, in agreement with result of Fig.~\ref{fig:dis,enu2bin}a (red histogram). For $E > 1$~TeV the minimum is at $\cos \theta_z < - 0.25$ (see blue histogram). For $\Delta m^2_{41} = 1 {\rm eV}^2$ and low energies the position of minimum is at $\approx - 0.02$, so that strong effect develops already in the first zenith angle bin. In the high energy range ($E > 1$ TeV) the minimum is at $\cos \theta_z < - 0.1$ in agreement with blue histogram of Fig.~\ref{fig:dis,enu2bin}b.

The break of the dependence at $\cos \theta_z \sim - 0.85$ corresponds to trajectories which start to cross the core. For $\cos \theta_z < - 0.85$ stronger suppression is due to the parametric enhancement of oscillations. The strongest relative effect is in the high energy bin which covers the resonance dip. For low energy bins the distortion is rather weak (distribution is almost flat), the $\nu_s$ effect is reduced to nearly uniform suppression of number of the $\nu_\mu$ events due to averaged oscillations. This can be absorbed by the uncertainty in the normalization of the atmospheric neutrino flux. Strong distortion comes from high energy bins where resonances are situated.

\begin{figure}[t!]
\centering
 \includegraphics[width=0.5\textwidth]{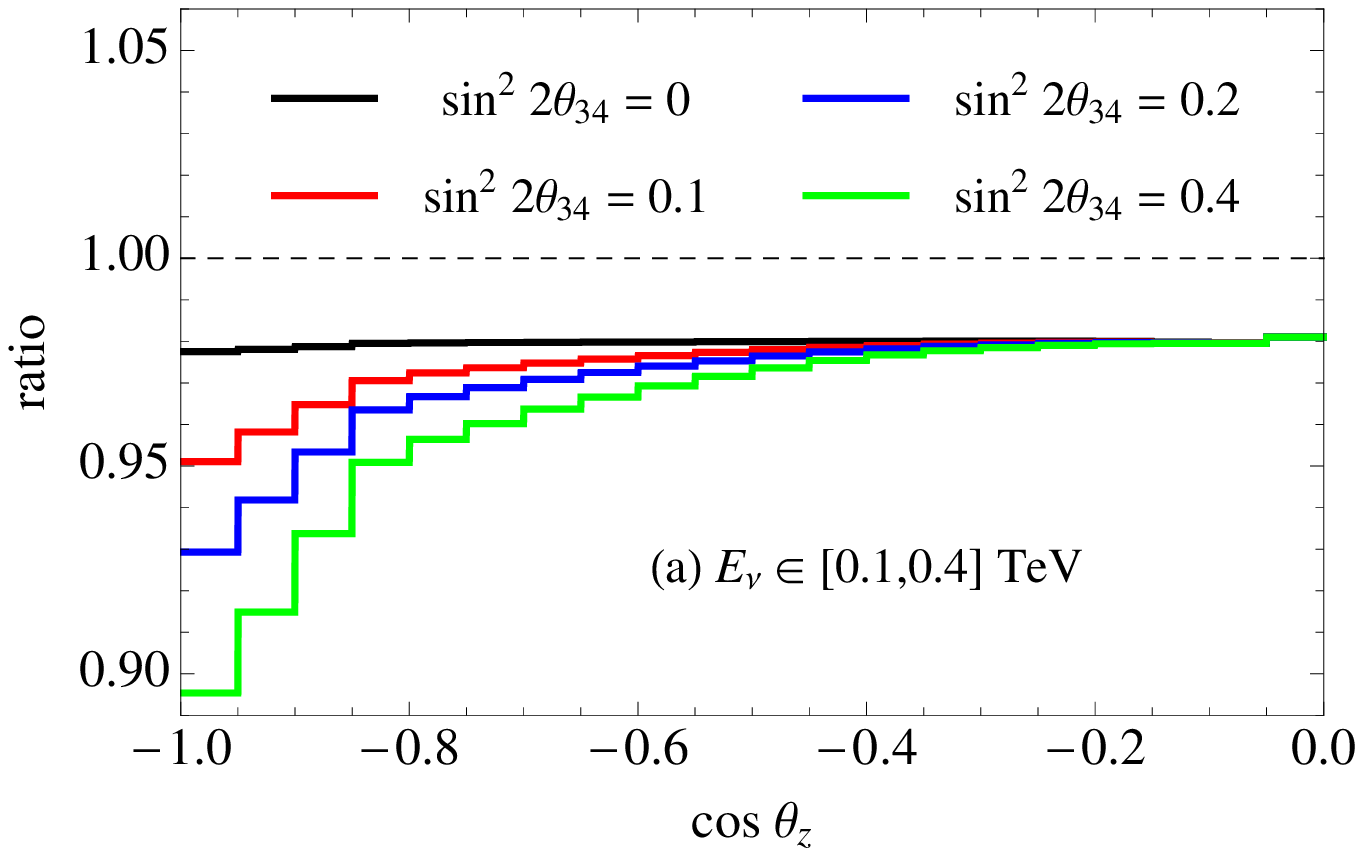}\includegraphics[width=0.5\textwidth]{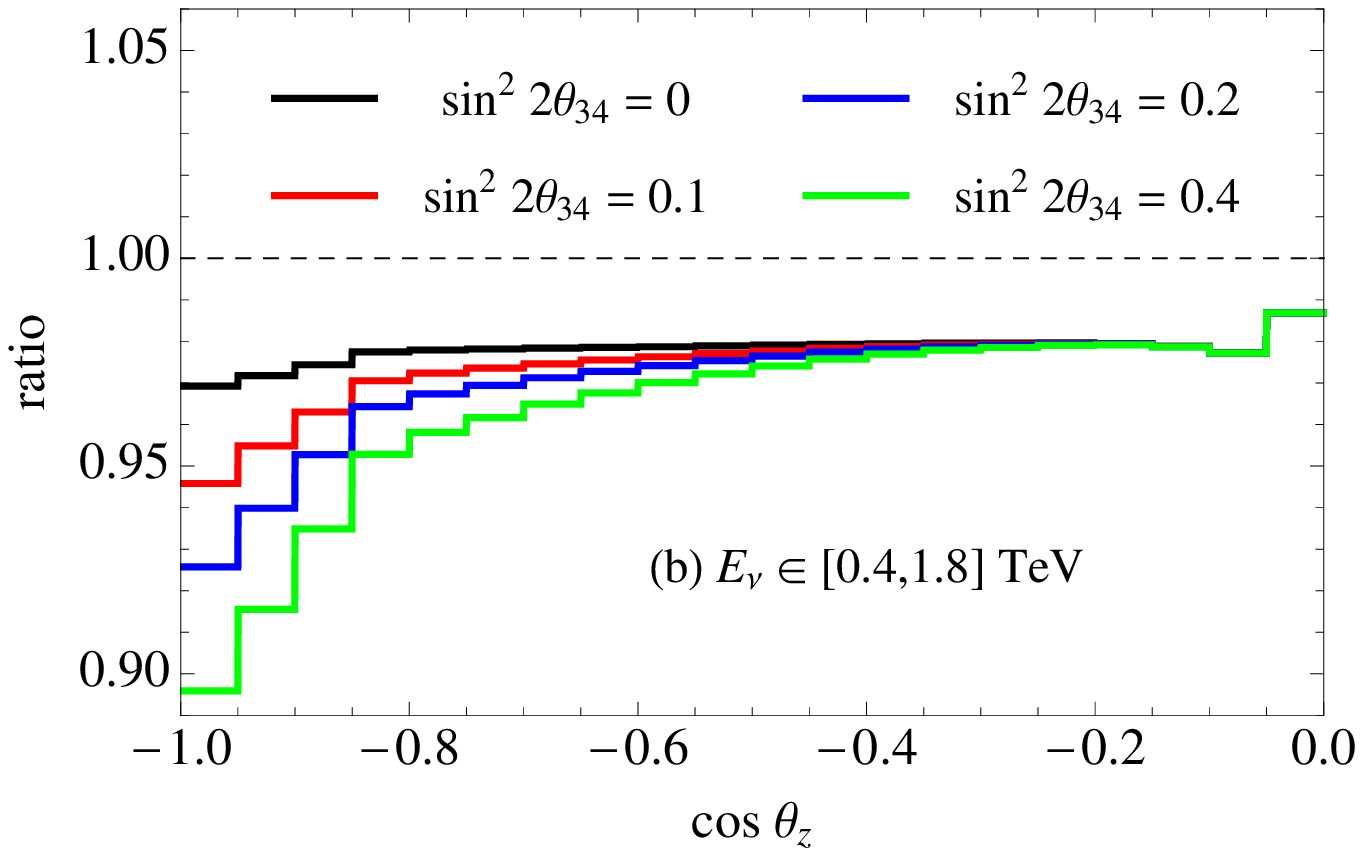}
\quad
 \includegraphics[width=0.5\textwidth]{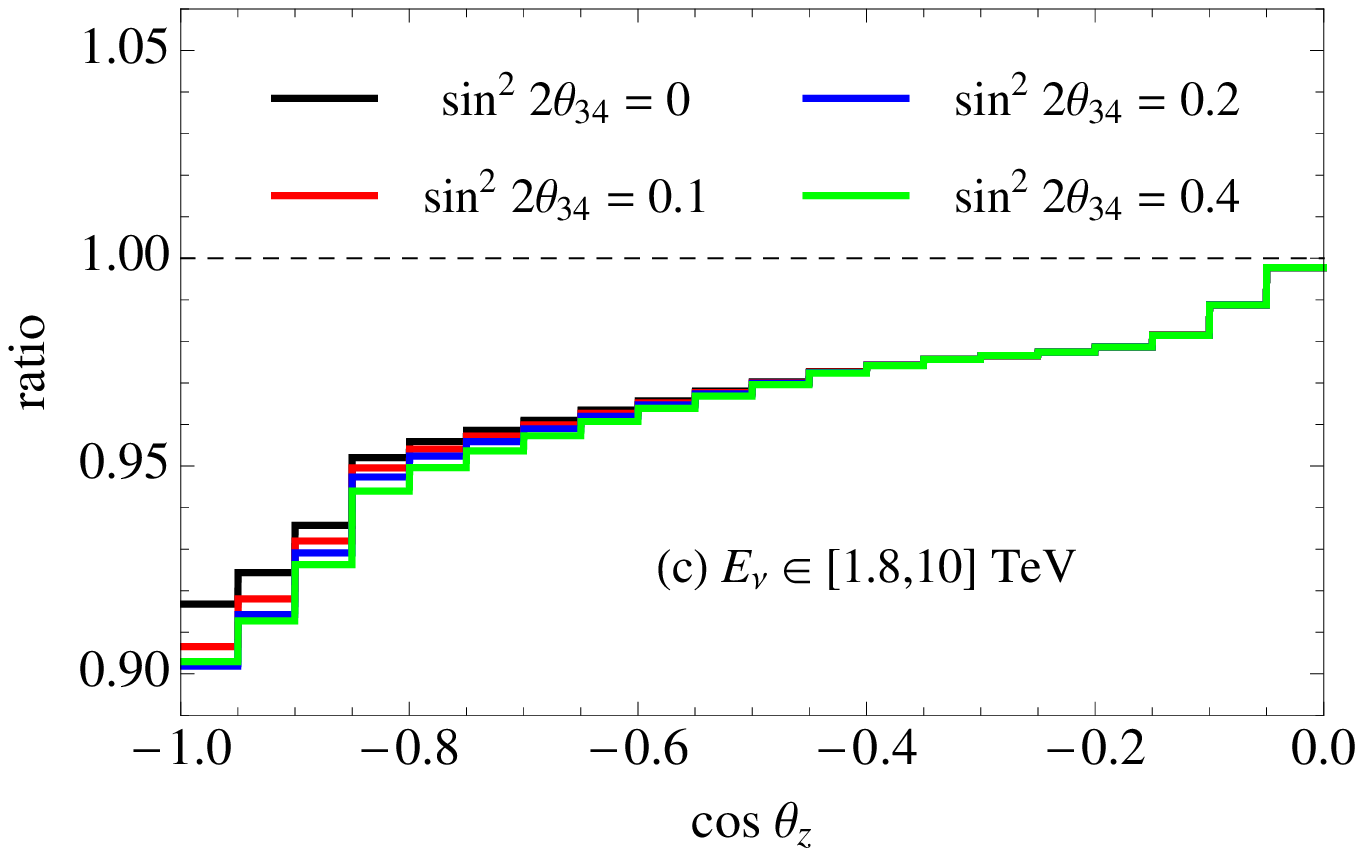}\includegraphics[width=0.5\textwidth]{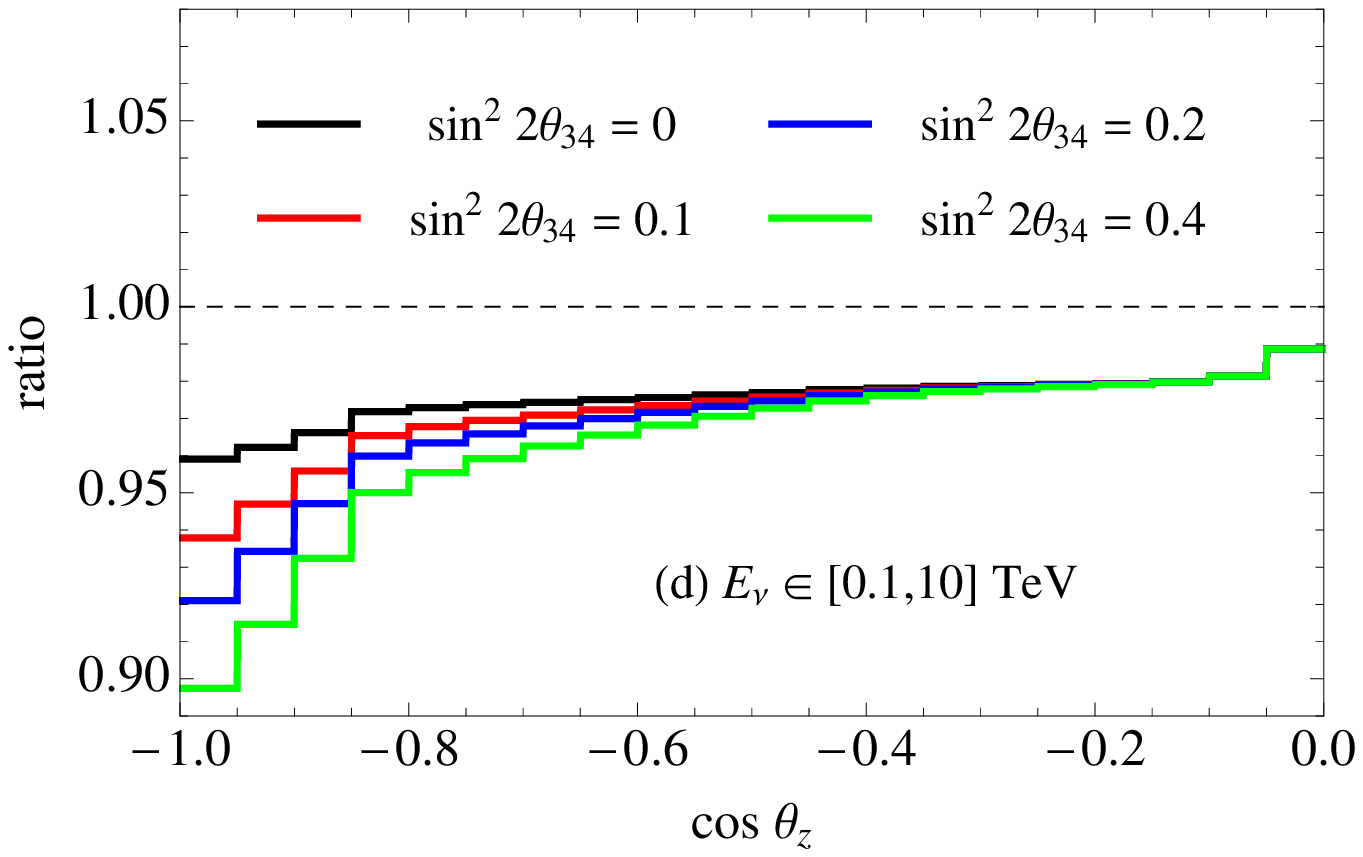}
\caption{\label{d}Distortion of the zenith angle distributions of events for different values of $\sin^2 2\theta_{34}$; for $\Delta m_{41}^2=1\,{\rm eV}^2$ and $\sin^22\theta_{24}=0.04$. The events are integrated over different neutrino energy bins: a) $(0.1-0.4)$~TeV, b) $(0.4-1.8)$~TeV, c) $(1.8-10)$~TeV and d) $(0.1- 10)$~TeV.} 
\end{figure}

\begin{figure}[t!]
\centering
 \includegraphics[width=0.5\textwidth]{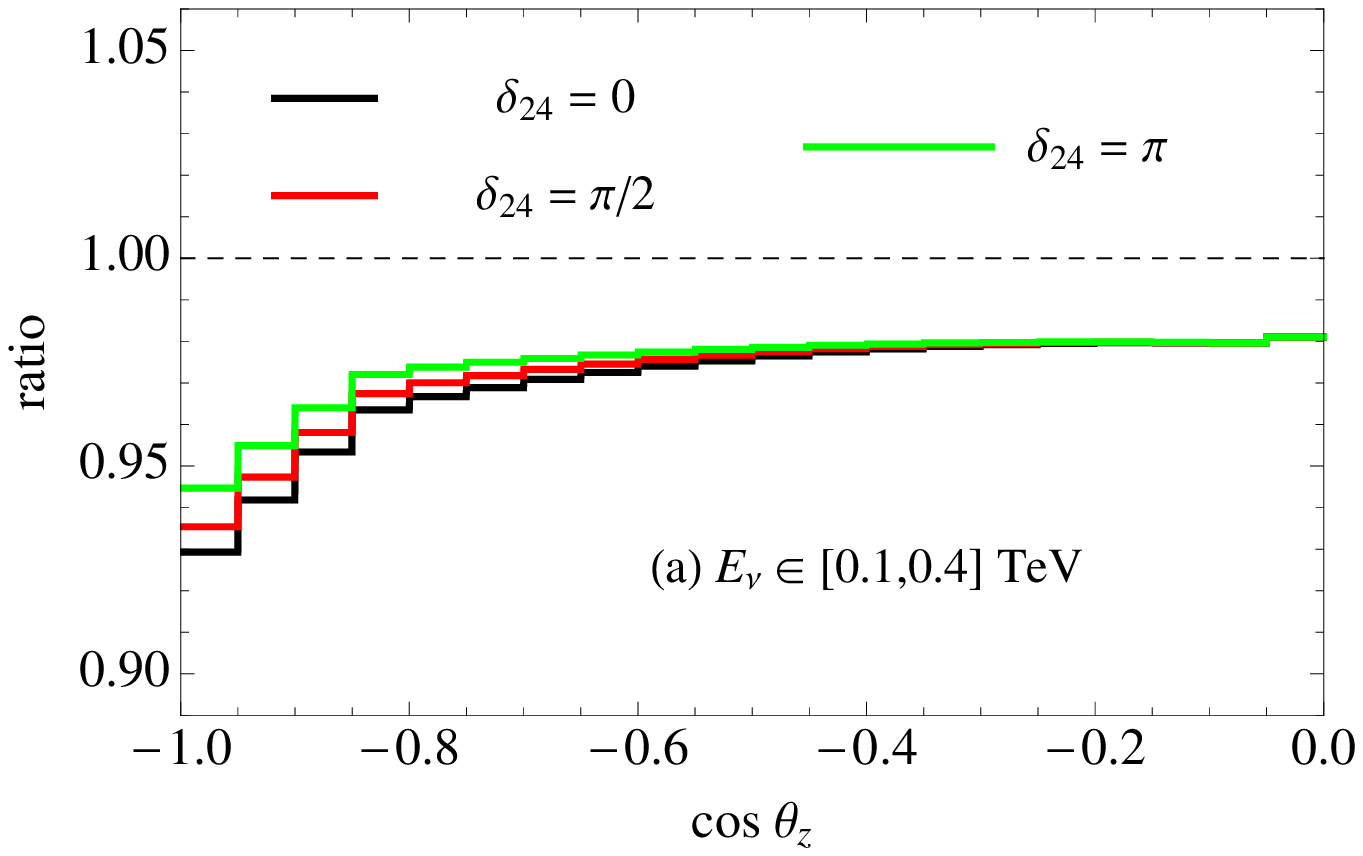}\includegraphics[width=0.5\textwidth]{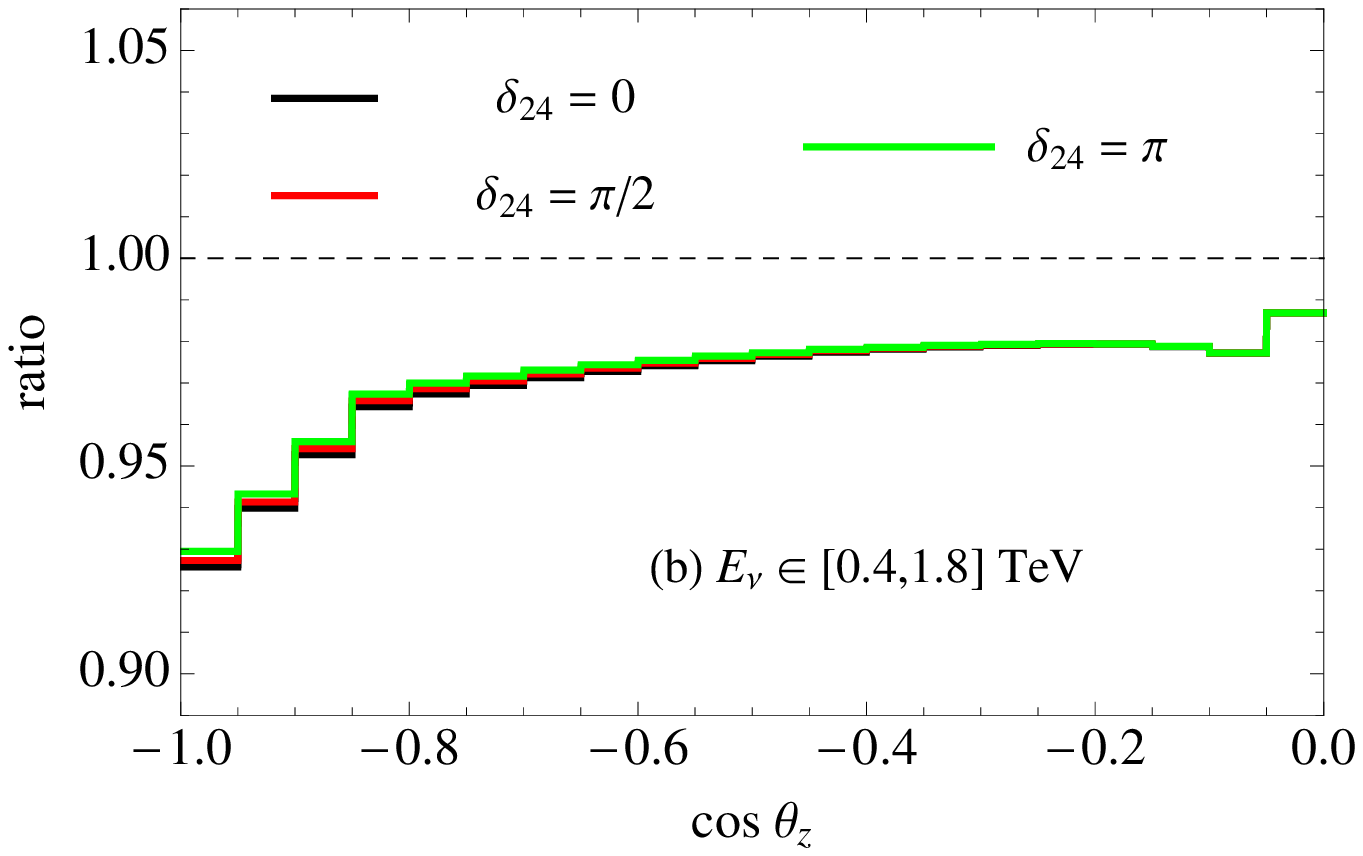}
\quad
 \includegraphics[width=0.5\textwidth]{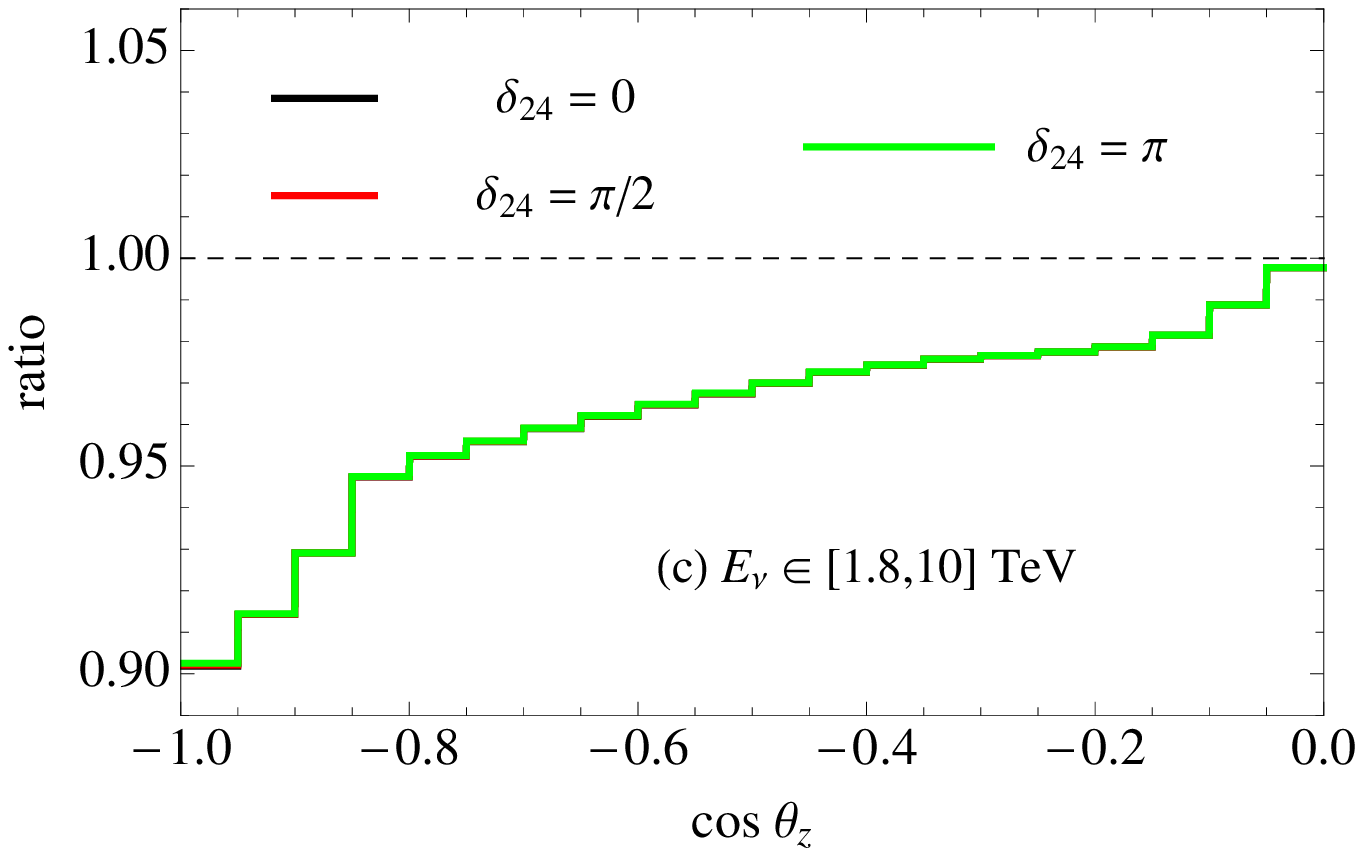}\includegraphics[width=0.5\textwidth]{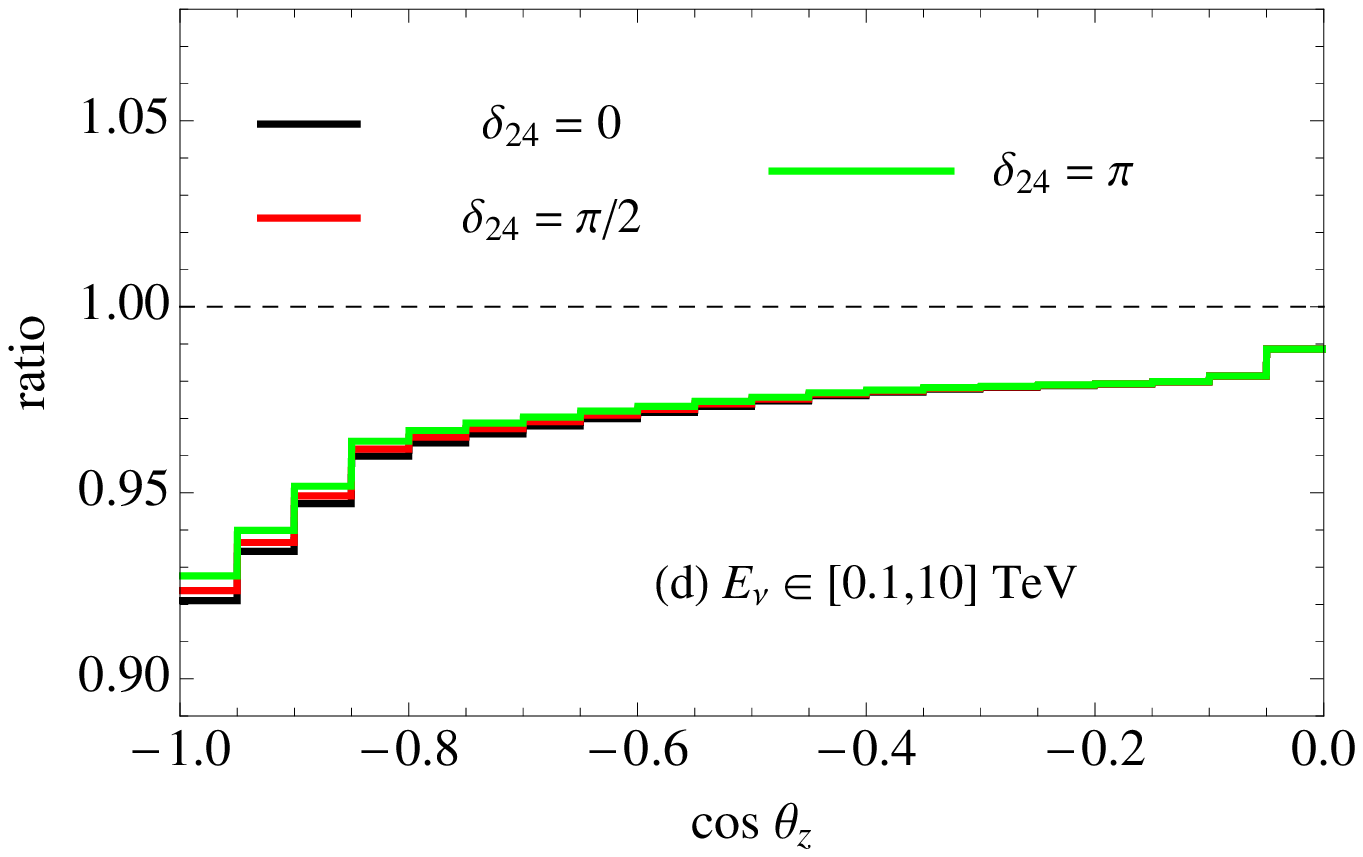}
\caption{\label{f}Distortion of the zenith angle distributions of events for different values of the CP-violating phase $\delta_{24}$. Here we set $\Delta m_{41}^2=1\,{\rm eV}^2$, $\sin^22\theta_{24}=0.04$ and $\sin^22\theta_{34}=0.2$. The events are integrated over different energy bins: a) $(0.1-0.4)$~TeV, b) $(0.4-1.8)$~TeV, c) $(1.8-10)$~TeV and d) $(0.1- 10)$~TeV.} 
\end{figure}

In Fig.~\ref{d} we show distortion of the zenith angle distributions for different values of $U_{\tau 4}$. With the increase of $U_{\tau 4}$ the effect of $\nu_s$ increases. Furthermore, the distribution (which was flat for $U_{\tau 4}= 0$) becomes more distorted even for low energies. The suppression for vertical direction can be about 2 times stronger in comparison with $U_{\tau 4}= 0$ case. For $U_{\tau 4}= U_{\mu 4}$ the effect is about $20 - 30 \%$. In the high energy bin the effect of $U_{\tau 4}$ is much weaker, due to the compensation of effects from the tail and the resonance dip. Notice that in the vertical bins for $E_\nu = (1.8 - 10)$~TeV the suppression decreases when $U_{\tau4}$ increases from $0.2$ to $0.4$ (Fig.~\ref{d}c). The reason for this opposite trend is that in the high energy bin the effect of dip dominates and dependence of suppression of signal on $U_{\tau4}$ follows the one of the dip. We found that this effect is slightly larger for $\Delta m^2_{41} = 0.5$~eV$^2$. However, the integral effects still decrease with $U_{\tau4}$.  

In Fig.~\ref{f} we illustrate dependence of the zenith angle distributions on the CP-violating phase $\delta_{24}$. The effect of CP-violation increases when $\delta_{24}$ changes from 0 to $\pi$. According to our consideration in Sec.~\ref{sec:cp} the strongest effect of CP-phase, about $20\%$, is in the low energy interval $E_\nu = (0.1 - 0.4)$~TeV. It decreases with energy: $7\%$ in the range $E_\nu = (0.4 - 1.8)$~TeV and $< 1\%$ in $E_\nu = (1.8 - 10)$~TeV. The integrated effect is about $10\%$.  

The effect of CP-violation on the zenith angle distributions is subleading with respect to the effect of $U_{\tau 4}$. This is the consequence of partial cancellation of the CP-violating contributions from the $\nu$ and $\bar{\nu}$ channels. With increase of $\delta_{24}$ sterile neutrino effect decreases in the $\nu$ channel and increases in the $\bar{\nu}$ channel almost by the same amount at the probability level. However, due to higher neutrino cross-section (and slightly larger flux) the contribution from neutrino channel is bigger than from the antineutrino one, and consequently, the total effect of sterile neutrinos decreases with the increase of $\delta_{24}$. Since CP effect is subleading and proportional to $U_{\tau 4}$, the smallest $\nu_s-$effect is still for $U_{\tau 4} \approx 0$. 

\begin{figure}[t!]
\centering
 \includegraphics[width=0.5\textwidth]{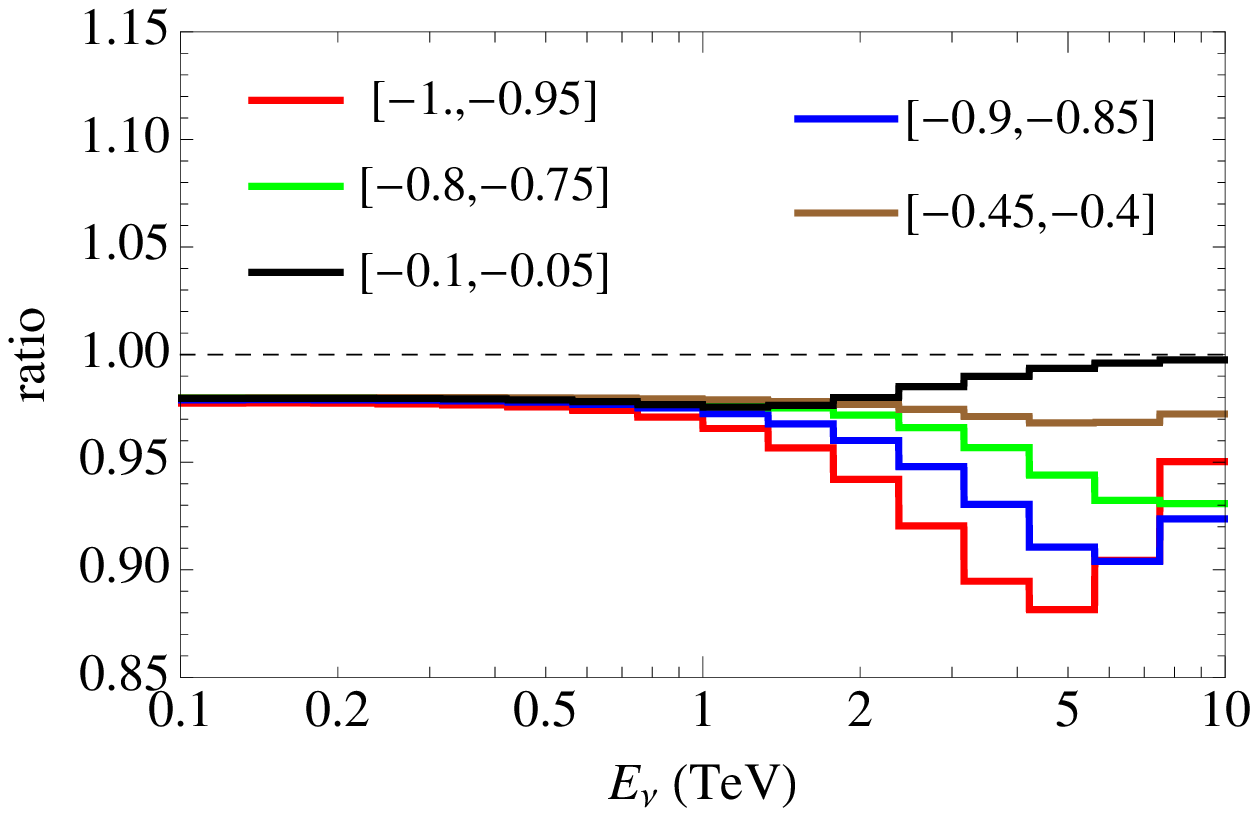}\includegraphics[width=0.5\textwidth]{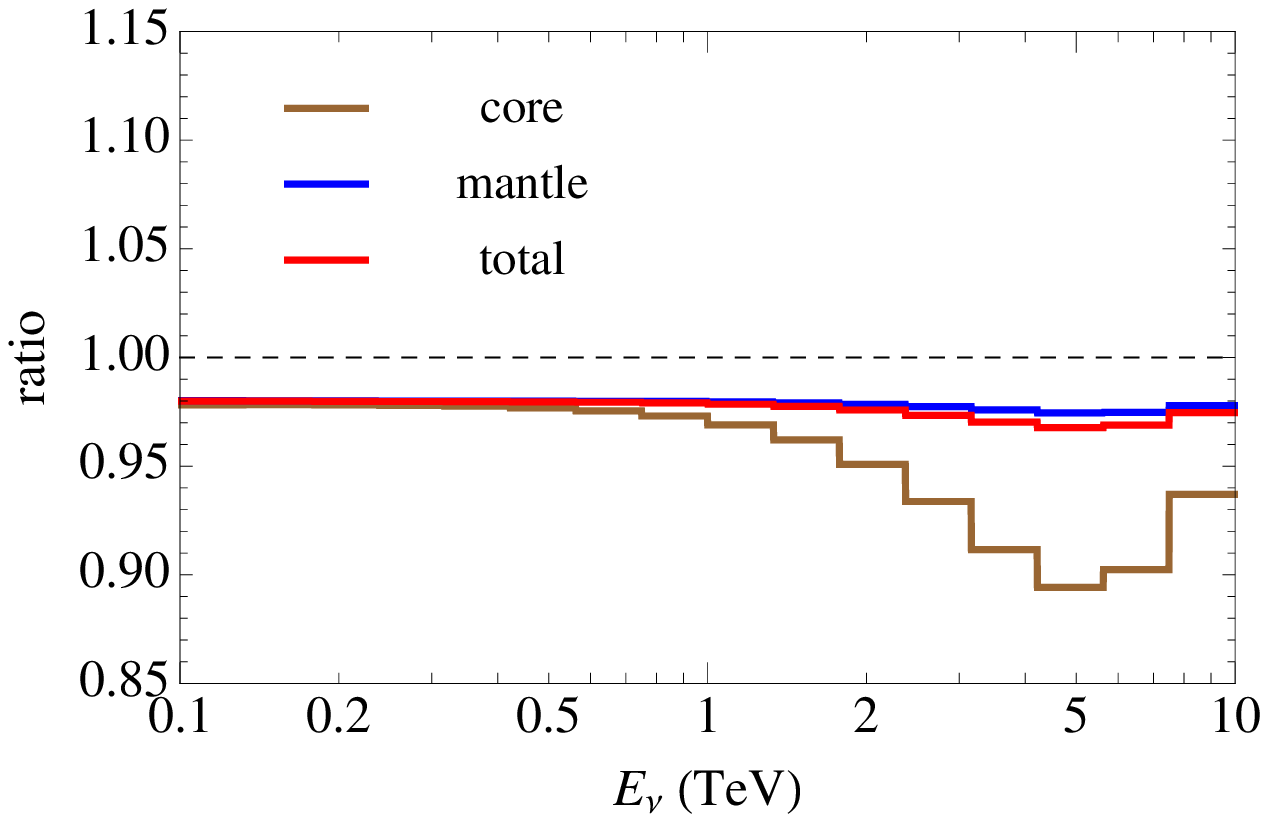}
\caption{\label{fig:dis,ener} Distortion of the smeared energy spectrum of the $\mu$-track events due to oscillations into sterile neutrinos for different intervals of the zenith angle. Smearing is performed with the neutrino energy reconstruction function $G$. Shown is the ratio of numbers of events with and without sterile neutrino mixing as function of the reconstructed energy. We used $\Delta m^2_{41}=1~{\rm eV}^2$, $\sin^2 2 \theta_{24} = 0.04$, $\theta_{34}=0$ and $\delta_{24} = 0$. Spectra are integrated over small intervals of $\cos \theta_z$ (left panel) and over trajectories which cross the core, the mantle and over all trajectories that cross the Earth (right panel).}
\end{figure}

For the better understanding of bounds on mixing of sterile neutrinos we present in Fig.~\ref{fig:dis,ener} distortion of the energy distribution of the $\mu$-track events due to oscillations into sterile neutrinos smeared with the reconstruction function $G(E_{\nu}^r, E_{\nu})$. Shown is the ratio of the number of events with and without sterile neutrino mixing as function of the reconstructed energy. We have performed integration of numbers of events according to Eq.~(\ref{eq:events,smear}) over small energy intervals, $\Delta E^r_\nu$, and different intervals of $\cos \theta_z$. In the left panel we show the ratios for energy spectra integrated over small intervals of $\cos \theta_z$. For the core-crossing trajectories the dip in the spectrum due to the parametric resonance is at $(5 - 6)$~TeV if  $\Delta m_{41}^2 = 1~{\rm eV}^2$. The dip in the survival probability is at 2.3~TeV. So, smearing of the probability multiplied by $A_{\rm eff} \Phi_{\nu_\mu}$ shifts the peak to higher energies by factor $\sim 2$. The shift is due to the decrease of $A_{\rm eff} \Phi_{\nu_\mu}$ with energy. As a consequence of the decrease, smearing produces stronger relative effect of the dip in probability on the distribution of events at higher energies. The shift factor equals $\sim2$ because the width of the smearing (reconstruction) function extends from very small energies to $2E_\nu$. For smaller values of $\Delta m_{41}^2$ the dip in the probability is at lower energies, where $A_{\rm eff} \Phi_{\nu_\mu}$ as a function of energy becomes flatter, and consequently, the shift becomes smaller.

For the mantle crossing trajectory with $\cos\theta_z=-0.8$ the dip due to the MSW resonance is at $E_\nu = 4$~TeV for $\Delta m_{41}^2 = 1$~eV$^2$ (see Fig.~\ref{fig:prob-antinu}). With increase of $\cos \theta_z$ the resonance energy slightly decreases. As in the core crossing case, the smearing shifts the dip by factor of two to $(8 - 10)$~TeV at $\cos \theta_z = - 0.8$. With increase of $\cos \theta_z$ the dip slightly shifts to lower energies, becomes shallow and then disappears. For trajectories close to horizon the oscillation effect disappears at high energies due to small baseline. At low energies the ratio in Fig.~\ref{fig:dis,ener} is given by the averaged vacuum oscillation probability $0.5 \sin^2 2\theta_{24}$. 

Results of integration over the core and the mantle crossing trajectories, as well as over all the trajectories are shown in the right panel of Fig.~\ref{fig:dis,ener}. The relative effect of sterile neutrinos is more profound for the core crossing trajectories. The integral mantle and core distributions have  the dips at the same energy $\sim 5$~TeV (for $\Delta m_{41}^2 = 1$ eV$^2$).

\section{Sensitivity of IceCube to sterile neutrino mixing}
\label{sec:sen}

In what follows we take $\theta_{14}=\theta_{34}=0$ (and vanishing CP-violation) which corresponds to the $\nu_\mu - \nu_s$ flavor mixing and the weakest effect of $\nu_s$ in IceCube. As we see from Figs.~\ref{fig:dis,enuint}, \ref{fig:dis,enu2bin}, \ref{fig:dis,enu3bin}, and the black curves of Figs.~\ref{d} and \ref{f}, sterile neutrinos produce distortion of the zenith angle distributions of the muon-track events which can be observed at IceCube. Even for the benchmark value of $\sin^2 2\theta_{24}=0.04$ which is much smaller that the values required by LSND and MiniBooNE and $|U_{\tau4}|^2 = 0$ the effect can reach $(5 - 6) \%$. At the same time uncertainties in the atmospheric neutrino flux, especially, the $\sim 24\%$ and $\sim 4\%$ uncertainties in the normalization and in the zenith dependence of neutrino flux (tilt)~\cite{Honda:2006qj}, respectively, can hide to some extend the distortion. Other smaller sources of uncertainty, such as pion to kaon ratio, are included effectively in the normalization uncertainty of 24\%. To estimate the sensitivity of IceCube to sterile neutrinos including these uncertainties, we introduce the following $\chi^2$ function:
\begin{equation}
\label{eq:chi2,2}
\chi^2(\Delta m_{41}^2,\theta_{24};\alpha,\beta)  = \sum_{i,j} \frac{\left\{N_{i,j}(\theta_{24}=0)-\alpha [1+\beta(0.5+(\cos\theta_z)_i)]  N_{i,j}(\theta_{24})\right\}^2}{\sigma_{i,j,{\rm stat}}^2+\sigma_{i,j,{\rm sys}}^2} +  \frac{(1-\alpha)^2}{\sigma_\alpha^2} + 
\frac{\beta^2}{\sigma_\beta^2}~, 
\end{equation} 
where $\alpha$ and $\beta$ are parameters which take into account the correlated uncertainties of the atmospheric  neutrino flux normalization and its zenith dependence (tilt) respectively. We use $\sigma_\alpha=0.24$ and $\sigma_\beta=0.04$~\cite{Honda:2006qj}. The tilt uncertainty implemented in Eq.~(\ref{eq:chi2,2}) in such a way that the zenith distribution of events, as in Figs.~\ref{fig:dis,enuint}-\ref{f}, can be rotated around the point $\cos\theta_z=-0.5$ with the angle determined by $\beta$. In Eq.~(\ref{eq:chi2,2}), $\sigma_{i,j,{\rm stat}}=\sqrt{N_{i,j}}$ is the statistical error and $\sigma_{i,j,{\rm sys}}=fN_{i,j}$ is the uncorrelated systematic error in $i$-th bin of $\cos\theta_z$ and $j$-th bin of $E_\nu^r$, with a parameter $f$ quantifying the systematic error. We will present the sensitivity of IceCube to sterile neutrinos assuming $f=0$, $5\%$  and $10\%$.  

\begin{figure}[t!]
\centering
\includegraphics[width=0.5\textwidth]{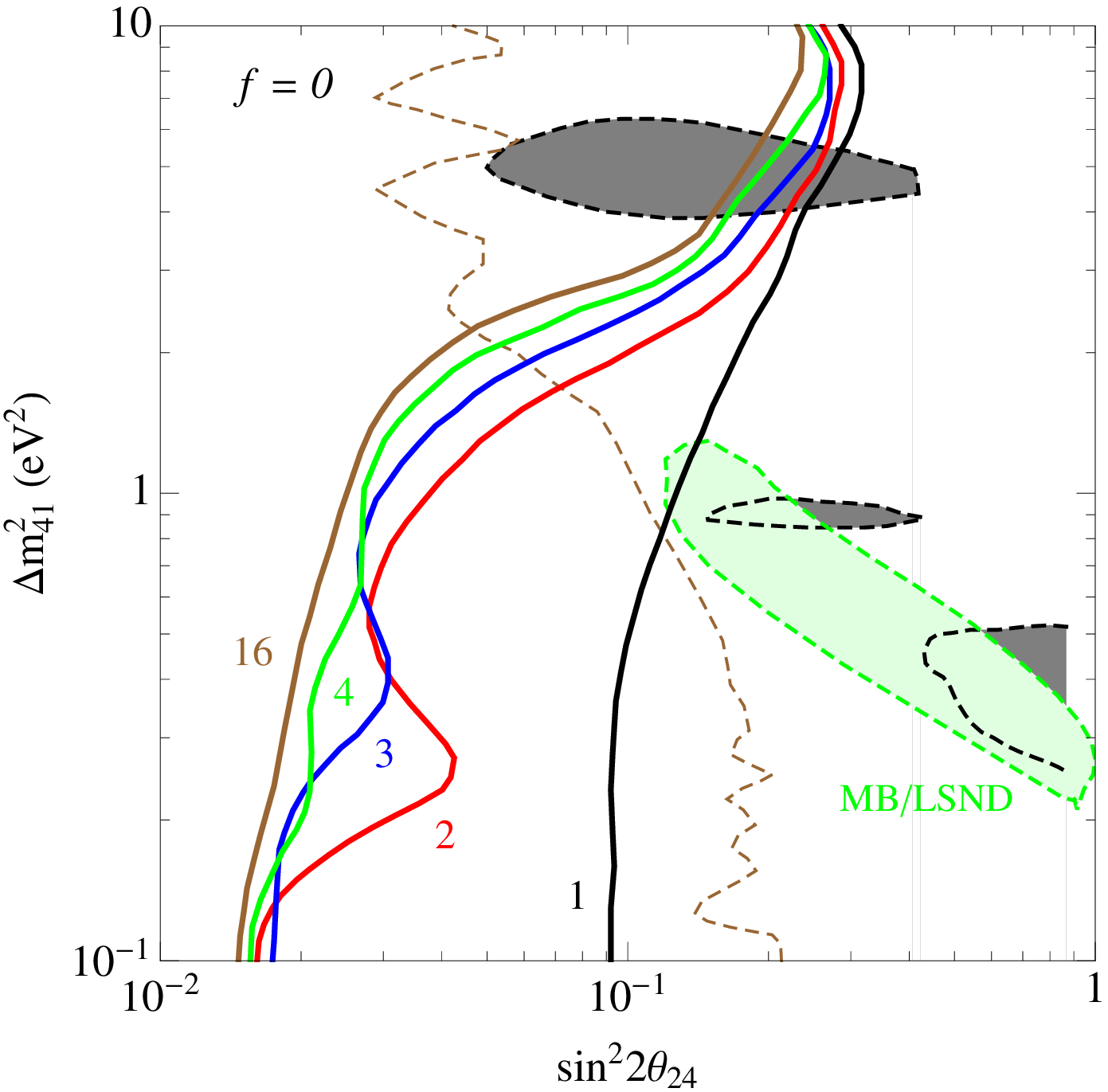}
\quad
\includegraphics[width=0.5\textwidth]{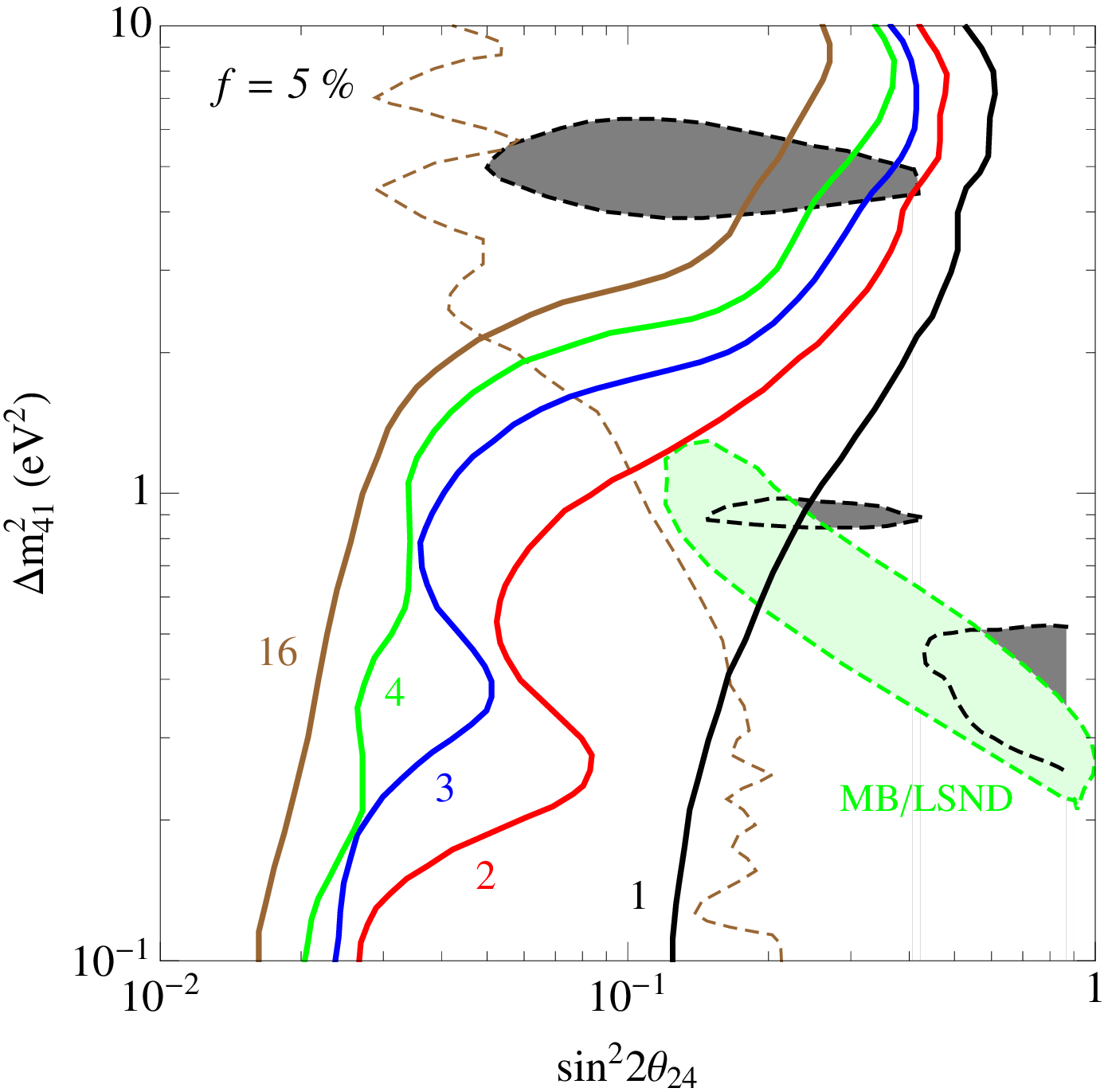}\includegraphics[width=0.5\textwidth]{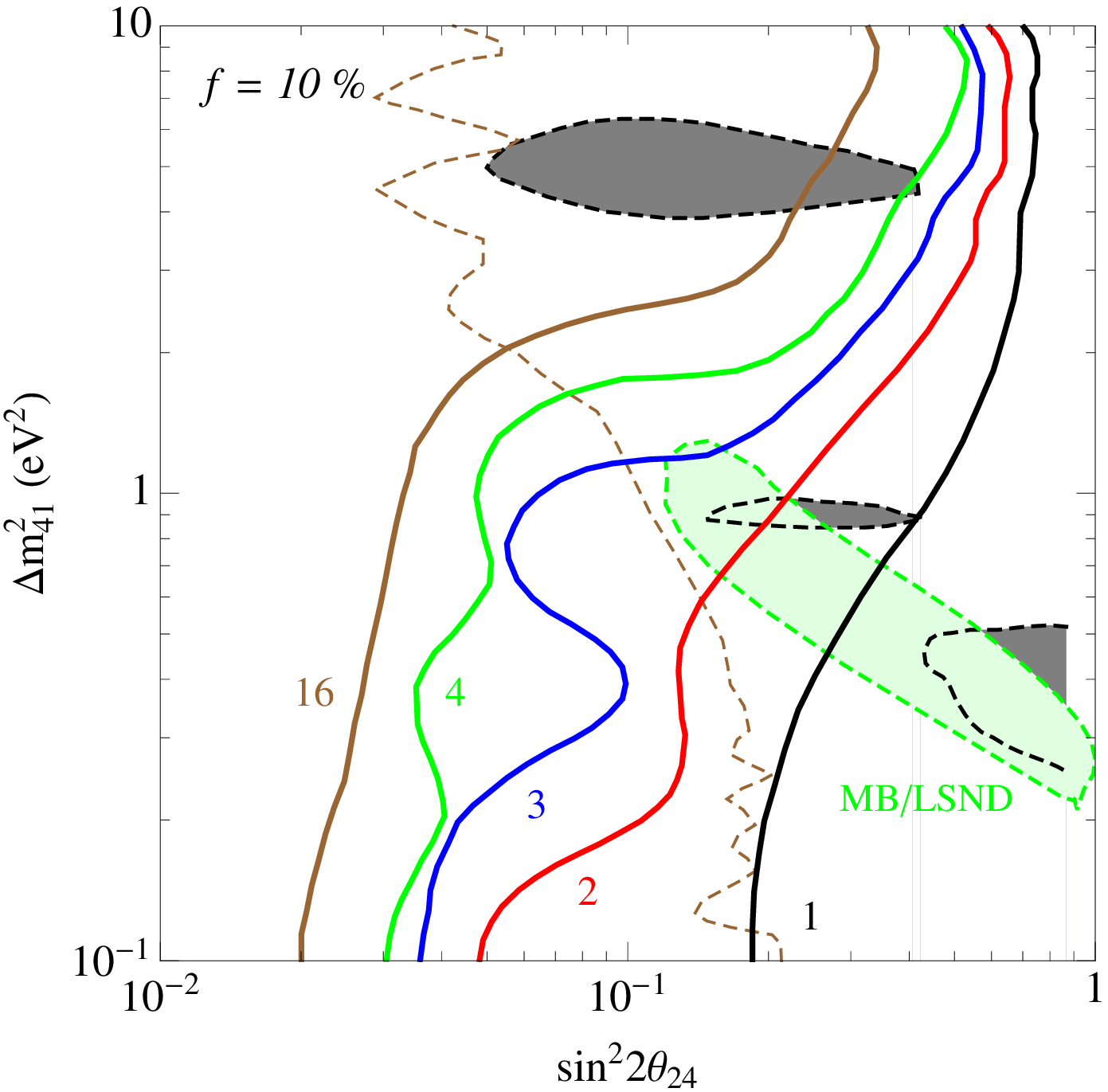}
\caption{\label{fig:limit,0sys}Sensitivity of IceCube to the 2-4 mixing of sterile neutrino at 99\% C.L. for different values of uncorrelated systematic errors and different energy binning. We take 3 times larger statistics than IceCube-79. The panels correspond to no error (top), $5\%$ (bottom left) and $10\%$ (bottom right) uncorrelated systematic errors. The numbers at the curves indicate the number of energy bins used in the analysis. The black dashed curves restrict the allowed region from combined analysis of the MiniBooNE, LSND, Reactor and Gallium data~\cite{Kopp:2013vaa}. The brown dashed curve shows the upper limit on $\sin^2 2 \theta_{24}$ from the combined $\nu_\mu$ disappearance data of MiniBooNE, CDHS and MINOS. The green dashed curve shows the allowed region from MiniBooNE and LSND data for the best-fit value $|U_{e4}|^2 = 0.023$~\cite{Kopp:2013vaa}.}
\end{figure}

In Eq.~(\ref{eq:chi2,2}) we essentially confront the number of events without sterile neutrino 
mixing which can be treated as the ``experimental'' value and the number of events with sterile neutrinos (fit). Fluctuations are not taken into account here. So, essentially $\chi^2$ evaluates ``distinguishability'' of the zenith angle distributions with and without sterile neutrinos. 

Three panels in Fig.~\ref{fig:limit,0sys} show the sensitivity of IceCube in the $(\sin^22\theta_{24},\Delta m_{41}^2)$ plane assuming uncorrelated systematic error $f=0$,  $5\%$ and $10\%$, respectively. The $99\%$ confidence level bounds have been obtained from the $\chi^2$ function in Eq.~(\ref{eq:chi2,2}). In the analysis we take 3 times larger statistics than IceCube-79 during 318 days has. We show dependence of the bounds on binning of events in the reconstructed energy: the curves correspond to no binning, 2 bins, 3 bins, 4 bins and 16 bins in energy in the range (0.1 - 10) TeV. 

According to the Fig.~\ref{fig:limit,0sys}, the sensitivity substantially improves with the energy binning. Two bins improve the bound by factor 5 at $\Delta m_{41}^2 \leq 1$~eV$^2$ in comparison with no-binning bound. In a sense this  binning  corresponds to reducing the uncertainties in the normalization and zenith dependence of neutrino flux. Further binning produces weaker improvement. In the case of small number of bins, 2 - 4 bins, the figures show local weakening of bounds at $\Delta m_{41}^2 \sim  (0.25 - 0.4)$~eV$^2$. This is an artifact of binning and the corresponding values of  $\Delta m_{41}^2$ depend on number of bins. Indeed, according to Fig.~\ref{fig:dis,ener} position of the resonance dip in the smeared distribution of events is at 
\be
E_{\rm dip}  \sim   5\; D(\Delta m^2_{41})~{\rm TeV} 
\left(\frac{\Delta m^2_{41}}{1~{\rm eV}^2}\right). 
\label{eq:dip}
\ee
Here $D(\Delta m^2_{41}) \sim 1$ for $\Delta m^2_{41} = 1 $ and it slightly decreases with the decrease of $\Delta m^2_{41}$. For two energy bins analysis (see Fig.~\ref{fig:limit,0sys}) the weakening of the bound is at $\Delta m^2_{41} = 0.23$ ${\rm eV}^2$. For this value of $\Delta m^2_{41}$ the position of the dip is at $E_{\rm dip} \sim 1~{\rm TeV}$, which coincides with the border between the bins. In statistical analysis splitting of a deviation (here from the no-oscillation case) into two bins reduces $\chi^2$, and consequently, weakens the bound\footnote{Indeed, if the deviation $d$ is in one bin, then $\chi^2 \sim d^2$, whereas equal splitting of the deviation into two bins will give $\chi^2 \sim 2(d/2)^2 = d^2/2$.}. In the case of 3 bins the weakening is at $\Delta m^2_{41} = 0.4~{\rm eV}^2$ which corresponds to $E_{\rm dip} \sim 2~{\rm TeV}$. This equals approximately the energy of border between the second and the third bins, $\sim 1.8$~TeV. In the case of 4 bins there are two regions of weakening of the bound: at $\Delta m^2_{41} = 0.2~{\rm eV}^2$ and $0.7~{\rm eV}^2$. The corresponding energies of the dip, 0.8~TeV and 3~TeV, coincide with the borders between the second and the third, and the third and the fourth bins.

Already for 4 bins the exclusion line becomes rather smooth. The curve corresponding to 16 bins of energy shows the maximum exclusion power of IceCube. Comparison of bounds in different panels of the Fig.~\ref{fig:limit,0sys} shows that increase of number of bins reduces effect of  systematic errors. The latter is big in the cases of  no binning or two bins. 

The overall bound weakens at high $\Delta m_{41}^2$. Indeed, for high $\Delta m_{41}^2$ the resonance region shifts to high energies and two factors contribute to this weakening: 1) $E_\nu A_{\rm eff} \Phi_{\nu_\mu}$ becomes small at high energies and therefore statistics decrease; 2) at high energies systematic uncertainties related in particular to flux normalization and tilt increase. Clearly better knowledge of the overall atmospheric neutrino flux normalization (that is smaller $\sigma_\alpha$) would help to derive stronger limits in this region. Notice that the IceCube and laboratory (accelerator) bounds are complementary: IceCube gives the strongest bound in the low $\Delta m_{41}^2$ range. 

According to Fig.~\ref{fig:limit,0sys},  for $\Delta m_{41}^2 < 2$ eV$^2$ the upper bound 
\be 
\sin^2 2 \theta_{24} < (3 - 4) \times 10^{-2} ~~~~~ (99\% ~{\rm C.L.})
\ee
or $|U_{\mu 4}|^2 < 10^{-2}$ can be established with already collected IceCube statistics. We find that the LSND and MiniBooNE  preferable range of parameters shown in Fig.~\ref{fig:limit,0sys} can be excluded at $(5 - 6)\sigma$ level. This result corresponds to the best-fit value $|U_{e4}|^2 = 0.023$~\cite{Kopp:2013vaa}. Larger allowed values of $U_{e4}$ will shift the green shaded region in Fig.~\ref{fig:limit,0sys} to the left and the level of exclusion reduces. We find that taking $|U_{e4}|^2=0.05$ at the upper 95\% C.L. allowed region will lead to $4.9 \sigma$ exclusion for $f = 0$, $4.5 \sigma$ for $f = 5\%$ and $3.9 \sigma$ for $f = 10\%$. A global fit of the IceCube, reactor and LSND/MiniBooNE data should be done to include uncertainty in $U_{e4}$ in more consistent way. Introduction of the second sterile neutrino would further strengthen the limit.

\section{Conclusions}
\label{sec:conc}

1. Tests of sterile neutrino interpretation of the LSND and MiniBooNE anomalies is mandatory in view of the fact that existence of eV scale neutrinos is not a small perturbation of the $3\nu$ picture both from theoretical and phenomenological points of views.

2. The mixing leads to disappearance of the $\nu_\mu$ ($\bar{\nu}_\mu$) atmospheric neutrino flux propagating through the matter of Earth at energies $E_\nu > 100$~GeV. The main effects are the resonance enhancement of oscillations in the antineutrino channel and the parametric enhancement of oscillations for the antineutrino  trajectories crossing the core of  Earth. Oscillations lead to specific distortion of the zenith angle and energy distributions of the $\mu$-track events in IceCube.

3. Due to matter effect the IceCube signal depends not only on $U_{\mu 4}$ but also on $U_{\tau 4}$ and new CP-phase $\delta_{24}$ and much weaker on other parameters. We explored dependence of the oscillation probability and number of events on $U_{\tau 4}$ and $\delta_{24}$. The dependence on $U_{\tau 4}$ is weak when $s_x^2 = (U_{\mu 4}^2 +  U_{\tau 4}^2)$ is small. It becomes substantial for $\phi_4 s_x \sim 1$. 

The effect of CP-violating phase $\delta_{24}$ on the probability is important for low energies and it is as large as $U_{\tau 4}^2$ effect. In fact, change of $\delta_{24}$ from $0$ to $\pi$ is similar to switching from neutrinos to antineutrinos. However, due to partial cancellation of the neutrino and antineutrino signals in the total signal, the CP-violation effect becomes subleading with respect to the $U_{\tau 4}^2$ effect.

4. We computed the zenith angle distributions of events in different energy ranges. Generic features of the distributions for small mixings are: (i) smooth increase of suppression with the decrease of $\cos \theta_z$, (ii) stronger suppression for the core crossing trajectories due to parametric enhancement of oscillations. The strongest effect is in the energy bin which covers the resonance dips in $P_{\mu\mu}$. The distributions change with $U_{\tau 4}^2$ and $\delta_{24}$ in monotonous and regular way for values of parameters considered in this paper. This allows us to conclude that the weakest effect of $\nu_s$ with $\Delta m_{41}^2 \sim  (0.5 - 1)$~eV$^2$ is realized for $U_{\tau 4} = 0$ and therefore $\delta_{24} = 0$. We also presented the energy distributions of the $\mu$-track events smeared with the neutrino energy reconstruction function. We showed that smearing leads to shift of the resonance dips to higher energies by factor $\sim2$.  

5. Using a simple $\chi^2$ method we have estimated sensitivity of the IceCube to the sterile neutrino mixing for $U_{\tau 4} = 0$. The sensitivity drastically improves when the energy information is included. The bounds become more stringent  with increase of number of the energy bins, partially because this reduces the role of systematic errors. We find that with the energy binning and 3 years exposure, which is already available now, IceCube can establish the bound $|U_{\mu 4}|^2 < 10^{-2}$ ($99\%$ CL) and exclude the region suggested by LSND and MiniBooNE  experiments with more than $(4 - 6)\sigma$ confidence level. 

The bounds we have obtained are subject to uncertainties related to the neutrino  energy  reconstruction, systematic errors, value of $U_{e4}$ and simplified statistical analysis. Improvements of the analysis would require the Monte Carlo simulations of events with sterile neutrinos. We expect, however, that these uncertainties will not reduce the confidence level of our bounds substantially.  

\section*{Acknowledgment}
A.~E. thanks ICTP for hospitality, where this work initiated and partially done. A.~E. acknowledges financial support by the funding grant 2009/17924-5, from S\~ao Paulo Research Foundation (FAPESP). 


 
\end{document}